\documentclass{aa}
\usepackage{graphicx}
\begin{document}
   \title{A Broadband Study of Galactic Dust Emission}


   \author{Paladini R.\inst{1},\inst{2} 
          Montier L.\inst{1},
          Giard M.\inst{1},
          Bernard J.P.\inst{1},
          Dame T.M.\inst{3},
          Ito S.\inst{4},
          Macias-Perez J.F.\inst{5}
          }

   \offprints{paladini@ipac.caltech.edu}

   \institute{Centre d'Etude Spatiale des Rayonnements,
           9, Avenue du Colonel Roche, Boite postale 4346, F-31028 Toulouse, France\\
           \and
            Spitzer Science Infrared Center, California Institute of Technology,
            1200 E. California Blvd, Pasadena, CA 91125, USA\\
            \email{paladini@cesr.fr,paladini@ipac.caltech.edu}
           \and
            Harvard-Smithsonian Center for Astrophysics, 
            60 Garden Street, MS 72, Cambridge, MA 02138, USA\\
            \and
            Department of Astrophysics, Nagoya University, 
            Furo-cho, Chikusa-ku, Nagoya 464-8602, Japan\\
            \and
             Laboratoire de Physique Subatomique et de Cosmologie, 53 
             Avenue des Martyrs, 38026 Grenoble Cedex, France\\
             }

   \date{}

   \abstract{We have combined infrared data with H$_{\sc{I}}$, H$_{\sc{2}}$ and H$_{\sc{II}}$ surveys  
in order to spatially decompose the observed dust emission into components associated 
with different phases of the gas. An inversion technique is applied. For the decomposition, we use the IRAS 
60 and 100 $\mu$m bands, the DIRBE 140 
and 240 $\mu$m bands, as well as Archeops 850 and 2096 $\mu$m wavelengths. In addition, we 
apply the decomposition to all five WMAP bands. We obtain longitude and 
latitude profiles for each wavelength and for each gas component in carefully selected Galactic radius bins. 
We also derive emissivity coefficients for dust in atomic, molecular and ionized gas in each of the bins. 
The H$_{\sc{I}}$ emissivity appears to 
decrease with increasing Galactic radius indicating that dust associated with atomic 
gas is heated by the ambient interstellar radiation field (ISRF). By contrast, we find evidence 
that dust mixed with molecular clouds is significantly heated by O/B stars still embedded in their 
progenitor clouds. 
By assuming a modified black-body with emissivity law $\lambda^{-1.5}$, we also derive the 
radial distribution of temperature for each phase of the gas. All of the WMAP bands except W appear to be dominated by emission from
something other than normal dust, most likely a mixture of thermal  
bremstrahlung from diffuse ionized gas, synchrotron emission and spinning dust.
Furthermore, we find indications of an emissivity
excess at long wavelengths ($\lambda \ge$ 850 $\mu$m) in the outer
Galaxy (R $>$ 8.9 kpc).  This suggests either the existence of a very cold
dust component in the outer Galaxy or a temperature dependence of the
spectral emissivity index.  Finally, it is shown that $\sim$ 80$\%$ of
the total FIR luminosity is produced by dust associated with atomic
hydrogen, in agreement with earlier findings by Sodroski et al.
(1997).\\
The work presented here has been carried out as part of the development of analysis tools
for the planned European Space Agency (ESA) Planck mission.


}

   \maketitle
%

\section{Introduction}

The release (Neugebauer et al. 1984) of the Infrared Astronomical Satellite  
(IRAS) maps (at 12, 25, 60 and 100 $\mu$m) provided 
a new perspective on the Galactic interstellar medium (ISM) by revealing 
diffuse infrared emission over almost the entire 
sky. This finding was confirmed (Hauser 1993) by the Diffuse Infrared Background 
Explorer (DIRBE) experiment (1.25 to 240 $\mu$m) on 
board the COBE satellite. Even in the lowest 12-$\mu$m IRAS band, stellar radiation
contributes only a small fraction ($\sim$ 8$\%$) of the emission (Boulanger $\&$ Perault 1988), the 
rest is from ISM. This fact is also corroborated by the 
tight correlations found by Boulanger et al. 1996 at high Galactic latitudes, i.e. $|b| >$ 20$^{\circ}$, between the 100 $\mu$m emission 
and the 21-cm emission, the latter being a faithful tracer of HI. By using a fixed spectral emissivity index $\beta$=2, 
these authors derive a temperature for dust associated with HI of 17 K.  
A similar correlation has been found for high-latitude molecular clouds, their CO emission being 
tightly correlated with the 100 $\mu$m intensity distribution (e.g. Boulanger et al. 1996). 
Likewise, Lagache et al. (2000) detect dust emission from the Warm Ionized Medium for $b <$ -30$^{\circ}$, $b 
>$ 25$^{\circ}$, with physical properties (temperature and emissivity) similar to dust associated with atomic 
hydrogen. Remarkably, most of this dust emission is not contributed by discrete H$_{\sc{II}}$ regions but by the diffuse 
gas.\\

As one can notice, all the results mentioned above have been obtained at high latitudes. In fact, the 
Galactic plane in in general much more complicated to investigate. This is because at high latitudes each 
line of sight traces a relatively short 
path through the Galaxy and this makes it possible to probe each distinct phase of the gas separately. 
On the contrary, in the Galactic plane the lines of sight are much 
longer and the phases of the gas heavily 
blended. A solution to this problem has been provided by the maximum-likelyhood method worked 
out by Bloemen et al. (1986). This method allows one to spatially decompose the total infrared 
emission into components associated with atomic, molecular and ionized gas in different bins of 
Galactic radiu. A similar technique is used in this paper and will be described in Section 2.\\

Another important observational fact derived from IRAS data is that most of the IR luminosity 
falls at $\lambda$ $>$ 60 $\mu$m. In particular, $\sim$ 50$\%$ of the IRAS luminosity is 
observed 
at $\lambda >$ 100 $\mu$m and only one sixth in each of the other bands. According to a variety of 
dust grain models (e.g. Mathis et al. 1977,
Draine $\&$ Lee 1984, Draine $\&$ Anderson 1985; Weiland et al. 1986; Desert et al. 1990; Li $\&$
Greenberg 1997; Dwek et al. 1997; Draine $\&$ Li 2001; Zubko, Dwek $\&$ Arendt 2004) the 12 $\mu$m emission is dominated by PAHs
(Polycyclic Aromatic
Hydrocarbons) which are characterized by a size of order of a few nanometers, while at 100 $\mu$m the emission is mainly due
to large grains, typically in the size range 10-20 nm to 0.1 $\mu$m. Intermediate wavelengths (25-60 
$\mu$m) are instead the domain of VSGs (Very Small Grains), in the nanometer range. Therefore, IRAS 
data indicate that most of the emission is undisputedly due to large grains. IRAS has also revealed 
a wide range of infrared dust colors from 12 to 100 $\mu$m 
(Boulanger et al. 1990). This can be explained through variations in
the dust size distribution. However, the same phenomenon has also been observed 
at submillimeter wavelengths (200, 260, 360 and 580 $\mu$m) by the PRONAOS balloon-borne 
experiment (Bernard et al. 1999, Stepnik et al. 2001) and it has been interpreted as due to 
changes in the physical properties of the 
large grains themselves, i.e. grain composition and temperature as well as optical properties.\\ 

The present work is motivated by several factors. In recent years there has been an 
unprecedented number of new centimeter and sub-millimeter experiments producing a large amount of data. 
Among these are the Archeops (Benoit et al. 2002) ballon-borne experiment 
and the WMAP ({\it{Wilkinson Microwave Anisotropy Probe}}) satellite 
(Bennett et al. 2003a). Such experiments have provided maps of the sky in the frequency range 23 to 545 
GHz (i.e. $\sim$ 10 mm down to 550 $\mu$m) at an angular resolution, depending on frequency, from $\sim$ 
50 down to $\sim$ 10 arcmin. Since most of the infrared emission is emitted at $\lambda >$ 60-100 
$\mu$m, 
the combination of these maps with existing IRAS and DIRBE maps now allows one to trace completely the 
dust spectrum in the frequency range where most of the emission is concentrated. At the same time, new 
all-sky data have also been released for the H$_{\sc{I}}$ emission (Section 3.1) and great progress has been made
in observing and modeling the ionized gas (Section 3.3). However, the most significant step forward has 
probably been in our understanding of the physics of dust grains. With respect to the color 
variations mentioned above, PRONAOS has also shed light on the dust emissivity 
index $\beta$: a significant 
inverse    
correlation between $\beta$ and dust temperature has been observed in the direction of star forming 
regions (Dupac et al. 2001). At the same time, analysis of the FIRAS data shows that, in the Galactic 
plane region, $\beta$ is best-fitted by a value of 1 (Reach et al. 1995) rather than 2 as usually 
assumed. These are very important findings that alone provide sufficient motivation for revisiting the problem 
of decomposing the Galactic infrared emission. Further motivation is provided by the impending Planck mission
(http//www.rssd.esa.int/~planck) which will obtain high-sensitivity, high-resolution ($\sim$ 5 to 30 arcmin)
all-sky maps in the wavelength range $\sim$ 400 $\mu$m - 10 mm. Due to its unique instrumental 
performance, Planck will provide unprecedented insights on the 
radio/infrared emission of the diffuse ISM of our Galaxy. The present work was carried out as part of a larger
effort to develop tools for the full exploitation of the Planck data. \\

The paper is organized as follows. Section 2 describes the inversion 
technique, including a brief 
review of previous works. In Section 3 we provide details regarding the datasets used, with particular 
attention given to the ionized gas. Section 4 presents 
our main results while Section 5 provides conclusions and perspectives.\\


\section{The inversion technique}

Along the Galactic plane, the detected infrared emission is a blend of 
radiation arising from dust that is spread over a wide range of distances and 
Galactic radii, and subject to a wide range of physical conditions. 
The purpose of the present study is to decompose this integrated infrared 
emission into radial bins associated with each phase of the interstellar gas and 
determine the physical properties of each phase in each bin. This goal can be achieved 
by means of a so-called {\em{inversion method}} that employs kinematic distances 
to assign gas to radial bins and require that the gas distributions in the adopted 
bins have distinctly different spatial distributions. 
The inversion technique has been first applied to astrophysical data by Bloemen et al. (1986) 
to determine the Galactocentric distribution of gamma-ray emissivity. 
Subsequently, Bloemen et al. (1990), Giard et al. (1994) and Sodroski et al. (1997) have 
used the same method to reconstruct the radial distribution of infrared emission. 
In the following, we will adopt the mathematical formalism introduced 
by these authors to describe in detail this technique.

Let $I_{\lambda}$ be the observed emission at wavelength $\lambda$ for a pixel $j$ of a map. Then the 
modeled emission $Im_{\lambda}$ for this pixel can be written as:

\begin{eqnarray}
Im_{\lambda}  & = & \sum_{i=1}^{n}\left ( \epsilon_{H_{\sc{I}}} (R_{i},\lambda) {N^{i}_{H_{\sc{I}}}} + \right. \\
              &   & {}  \left. \epsilon_{H_{\sc{2}}} (R_{i},\lambda){N^{i}_{H_{\sc{2}}}} + \epsilon_{H_{\sc{II}}}
 (R_{i},\lambda){N^{i}_{H_{\sc{II}}}}\right)\nonumber
\end{eqnarray}

\noindent
In the expression above: $i$ denotes the intervals (or rings) of galactocentric radii $R_{i}$ over which 
the decomposition 
is performed; $\epsilon_{H_{\sc{I}}} (R_{i},\lambda)$, $\epsilon_{H_{\sc{2}}}(R_{i},\lambda)$, $\epsilon_{H_{\sc{II}}} (R_{i},\lambda)$ 
are the dust emissivities associated with the different phases of the gas in each ring; 
${N^{i}_{H_{\sc{I}}}}$ represents the H atom column density for neutral atomic hydrogen (H$_{I}$) in the 
interval $R_{i}$; similarly, ${N^{i}_{H_{\sc{2}}}}$ is the H atom column density for molecular hydrogen (H$_{2}$) and 
${N^{i}_{H_{\sc{II}}}}$ is the H atom column density for ionized hydrogen (H$_{II}$) in the considered ring. 

Note that eq.~(1) does not require resolution of the kinematic distance ambiguity for material
within the solar circle, since we assume that the dust properties vary only with Galactic radius.
In addition, 
following Giard et al. (1994) and Sodroski et al. (1997) we do not assume a constant ratio 
between $\epsilon_{H_{\sc{I}}} (R_{i},\lambda)$ and $\epsilon_{H_{\sc{2}}}(R_{i},\lambda)$, rather these parameters 
are allowed to vary in each galactocentric ring. The dust emissivities can be determined 
by means of a least-square fit analysis, i.e., by minimization of the quantity:

\begin{equation}
\chi^2 = \sum_{j=pixels} \frac{({Im(j)}_{\lambda}-{I(j)}_{\lambda})^2}{{\sigma_{j}}^2} 
\end{equation}

\noindent
over the map. In the expression above, $\sigma_{j}$ ($\sigma_{j}$ = $\sigma_{j}$($\lambda$)) represents the noise per pixel for  
the I$_{\lambda}$ map (see Section 4.1 for details).

\noindent	
Column densities are computed, for every ring and for 
each phase of the gas, by applying a circular rotation model for the Galaxy. The rotational velocity 
$\theta(R)$ of a gas element in differential motion with respect 
to the 
Galactic center is given by:

\begin{eqnarray}
\theta(R) = \frac{R}{R_0} \left ( \theta_{0} + \frac{V_{LSR}}{\sin{l} \hskip 0.1 truecm \cos{b}} \right ) \hskip 2truecm [km/s]
\end{eqnarray}

\noindent
with $R$ the galactocentric distance of the considered gas element, $R_{0}$ the galactocentric 
distance 
of the Sun (we take $R_{0}$ = 8.5 kpc), $\theta_{0}$ the rotational velocity of the Sun 
(we assume $\theta_{0}$ $\simeq$ 220 km/s), $V_{LSR}$ the linear velocity along the line of sight in 
the Local 
Standard of Rest and $l$ and $b$ the Galactic coordinates for a given direction. We adopt the 
rotation curve $\theta(R)$ by Fich, Blitz $\&$ Stark (1989) for which the galactocentric 
distance can be expressed as:

\begin{equation}
R  = \left ( \frac{221.641\times R_{0}}{\theta_{0}+0.44286\times R_{0}+\frac{V_{LSR}}{\sin{l} \hskip 0.1  
truecm \cos{b}}} \right )  \hskip 1.2truecm [kpc]
\end{equation}

\noindent
Therefore, given a specified ring $R_{i}$, the 
corresponding range in linear velocities is:

\begin{eqnarray}
V^{i}_{LSR} & = & 221.641 \times \frac{R_{0}}{R_{i}} + \hskip 2.4truecm [km/s] \\
                  &  & {}  - \sin{l}\cos{l} (\theta_{0} + 0.44286\times R_{0}) \nonumber  
\end{eqnarray}

\noindent
For each given ring, ${N^{i}_{H_{\sc{I}}}}$,  ${N^{i}_{H_{\sc{2}}}}$ and ${N^{i}_{H_{\sc{II}}}}$ are then computed by integration over the velocity 
channels selected as above. Further details, specific to each gas phase, will be given in the following 
section. We note that the use of a different rotation model would not significantly affect 
the results of the analysis. For the ring decomposition, we have chosen the following 
intervals:

\begin{eqnarray}
R_{i} & = & \left [\hskip 0.1truecm [0.1, 4], [4, 5.6], [5.6, 7.2], [7.2, 8.9], \right. \hskip 0.5truecm [kpc]\\
          &  & {} \left. [8.9,14], [14,17] \hskip 0.1 truecm \right ] \nonumber
\end{eqnarray}

\noindent
The above decomposition has been obtained through a process of optimization of the 
number and type of Galactocentric intervals aimed at reducing 
the cross-correlation between neighboring rings and therefore leading to a well-defined 
linear inversion problem. This goal has been achieved by starting with a fine decomposition 
(roughly 200 bins in Galactic radius) and then gradually combining correlated rings, guided by the 
eigenvalues of the cross-correlation matrix for different components. 

\noindent
All the maps used for the inversion have been projected into HEALPix format (Gorski et al. 1999) 
and degraded to 1$^{\circ}$ resolution. The final maps (input IR maps as well as H$_{\sc{I}}$, H$_{\sc{2}}$ and H$_{\sc{II}}$ 
galactocentric-annuli maps) are at the HEALPix resolution (i.e. {\em{nside}}) 256 which roughly 
corresponds to a pixel linear size of 13.7$^{\prime}$.

\section{The data base}

\subsection{Dust}

To trace dust emission we use a large data base ranging from sub-millimeter to 
centimeter wavelengths. At short wavelengths, i.e. 60 and 100 $\mu$m, we use the last 
generation of IRAS plates, the so-called IRIS 
({\it{Improved Reprocessing of the IRAS Survey}}) 
maps (Miville-Deschenes $\&$ Lagache, 2005). These maps have been obtained by 
reprocessing the original ISSA plates in a manner that allowed the correction for 
calibration, zero level and striping problems. The IRIS maps (which are at an angular resolution of 4-arcmin at 60 $\mu$m 
and 4.3-arcmin at 100 $\mu$m) also have 
a better zodiacal light subtraction. At 140 and 240 $\mu$m, 
we use the DIRBE {\it{Zodi-Subtracted Mission Average Maps}}. DIRBE is 
a photometer on-board the COBE satellite with a resolution of  
40-arcmin. To probe the sub-millimeter wavelengths, we make use of 
Archeops data. Archeops is a ballon-borne experiment (Benoit et al. 2002) which has 
detected Cosmic Microwave Background (CMB) anisotropies at the average angular resolution of 
8-arcmin in 4 frequency bands, 143, 217, 353 and 545 GHz. It is the only experiment 
of this kind which has provided information on the Galactic plane emission at millimeter 
wavelengths over a large fraction of the sky. Three flights have been performed: a first one (T99) 
in July 1999 from Trapani (Italy) followed by a second one (KS1) in January 2001 from 
Kiruna (Sweden) and by a third one (KS3) in February 2002 again from Kiruna. For the 
present analysis, we use the T99-KS3 combined data at 143 and 353 GHz. These combined maps 
provide better sky coverage than individual maps. In particular, the T99-KS3 data set 
covers the longitude interval $\sim$ 30 $<$ l $<$ 180 degrees. At centimeter wavelengths, we use 
the WMAP all-sky maps. WMAP is a satellite dedicated to the observations 
of CMB anisotropies with a frequency coverage in the range 23 to 94 GHz and an angular 
resolution increasing with frequency from 52.8 to 13.2 arcmin.

\subsection{Atomic hydrogen}

\noindent
For the atomic hydrogen, we use the newly-released LAB 
(Leiden/Argentine/Bonn) survey (Kalberla et al. 2005). This data set has been obtained 
by merging the Leiden/Dwingeloo Survey (hereafter LDS, Hartmann \& 
Burton, 1997) of the 
sky north of $\delta$ = -30$^{\circ}$ with the Instituto Argentino de 
Radioastronomia Survey (hereafter IAR, Arnal et al. 2000; Bakaka et al. 
2005) which 
covers declinations south of $\delta$ = -25$^{\circ}$. The angular 
resolution of the final data cube is 0.6$^{\circ}$ while the velocity 
resolution is 1.3 km s$^{-1}$ with a LSR velocity range from -450 km 
s$^{-1}$ to +400 km s$^{-1}$. The IAR survey was specifically 
designed to complement the LDS survey of the northern sky. 
Although the observational parameters of the two surveys 
are quite similar, some reprocessing has been necessary to combine them. 
Both data sets have been corrected for stray 
light radiation. In addition, the existing LDS cube has been reprocessed 
to correct for reflected ground radiation. Owing mainly to limited computing
power, the original processing was not able to apply this correction accurately.
Consistency checks between the two surveys have 
also been made in the region of overlay (-30$^{\circ} < \delta <$ -25$^{\circ}$),
and they appear to agree quite well (Bajaja et al. 2005). 
To merge the two data sets into a single cube a spline interpolation followed 
by a spatial regridding has been performed. Such regridding has slightly 
degraded the original resolution of the two surveys to 37.5'. The rms 
brightness-temperature noise of the merged database is 0.07-0.09 K. 
Residual errors due to defects in the correction for stray radiation are mainly below 20-40 mK.\\

This data base is used to compute, for every ring $R_{i}$ and every given line of sight 
$k$, the 
quantity $(N_{H_{I}})^{i}_{k}$ 
according to (Kerr 1968):

\begin{eqnarray}
(N^{i}_{H_{I}})_{k} & = & 1.82 \times 10^{-2} T_{s} \times \\
                              &  & \int_{V^{i}_{LSR}} {\ln \left (1-\left(\frac{T_{b(k,v)}}{T_{s}}\right) \right
)^{-1}} dv \nonumber
\end{eqnarray}

\noindent
In the expression above, $T_{s}$ is the spin temperature ($T_{s}$=145 K) while 
$T_{b}$, in K, is the H$_{\sc{I}}$ observed brightness temperature. The integral is 
taken over the velocity range defined in eq.~(5). Units are 10$^{20}$ atoms cm$^{-2}$. 
It is important to emphasize that different assumed values 
of $T_{s}$ in the range 120-160 K would change computed column densities by
less than a few percent (Sodroski et al. 1997).

\begin{figure*}
\centering
\includegraphics[width=18cm,height=16cm,angle=90]{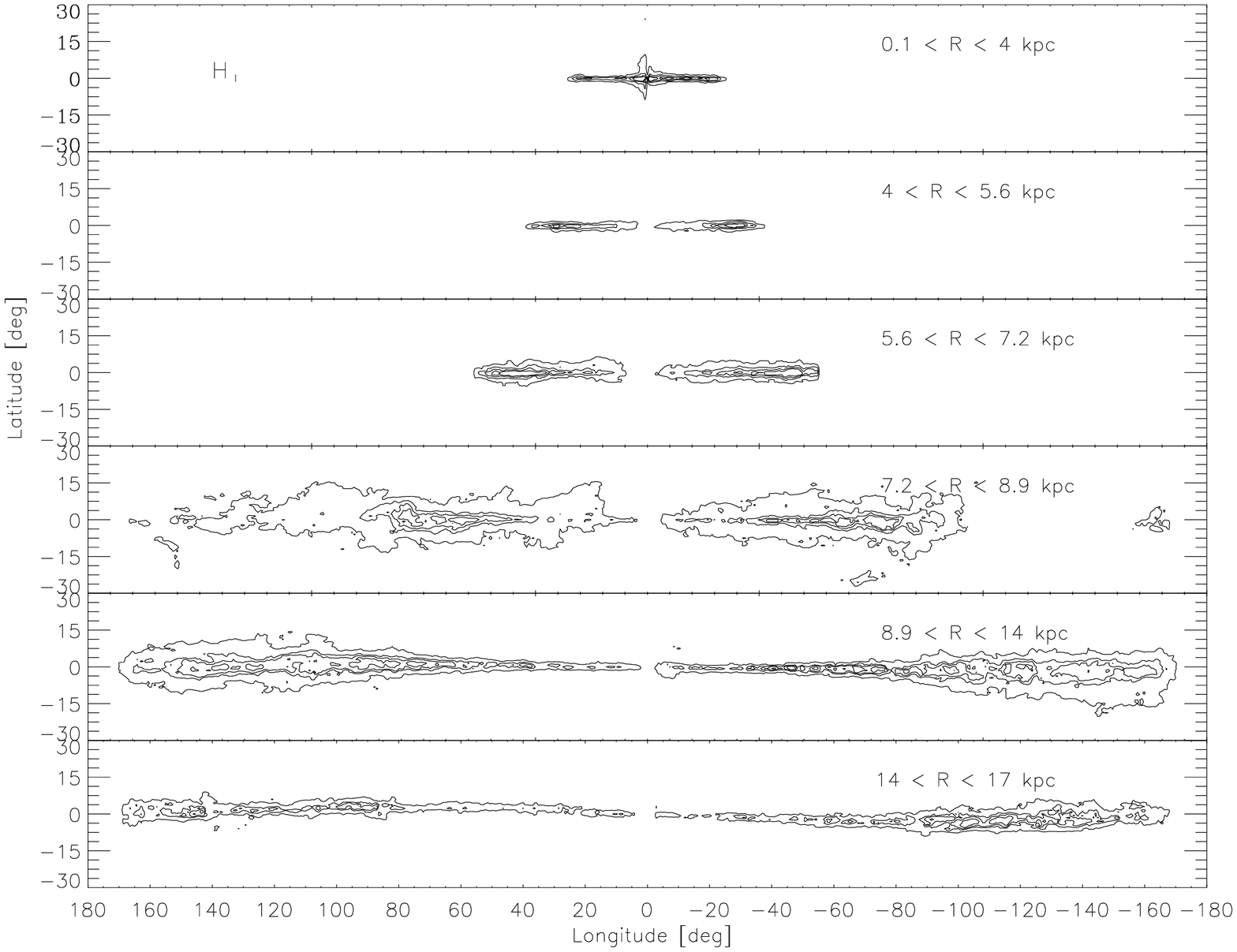}
\vspace*{0.3truecm}
\\
\caption{Ring 1 to 6 (from top to bottom) for H$_{{\sc{I}}}$ 
column densities. Levels of the contours are: 3, 30, 70, 110 and 150 10$^{20}$ atoms cm$^{-2}$. 
}
\end{figure*}

\begin{figure*}
\centering
\includegraphics[width=18cm,height=16cm,angle=90]{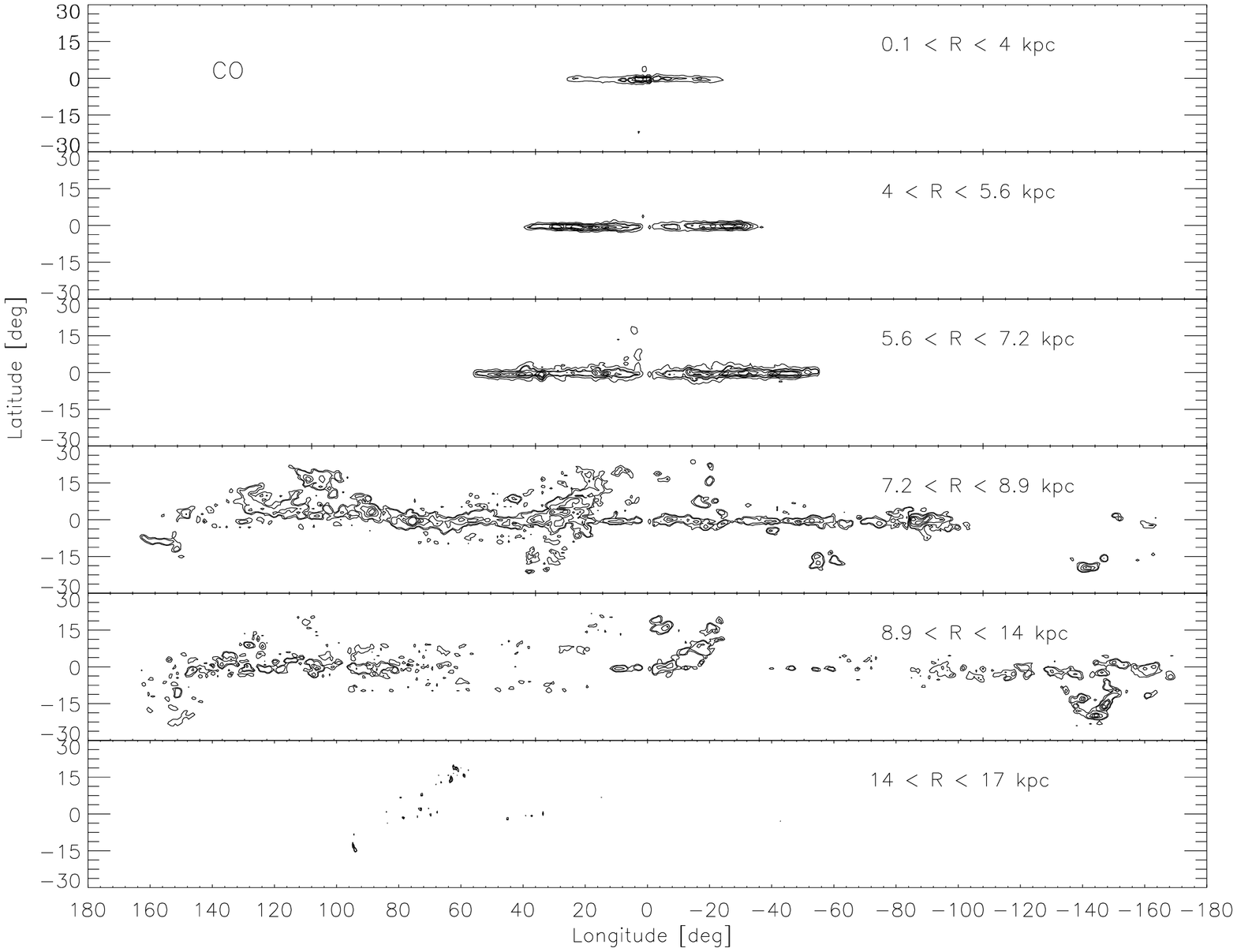}
\vspace*{0.3truecm}
\\
\caption{Ring 1 to 6 (from top to bottom) for CO 
column densities. Levels of the contours are: 
20, 40, 50, 90, 130 and 170 10$^{20}$ atoms cm$^{-2}$.
}
\end{figure*}

\noindent

Fig.~1 shows spatial maps of HI integrated over each of the radial bins. 
The emission has the widest latitude extent in the 7.2 $<$ R $<$ 8.9 kpc bin -- which includes 
the Sun -- and becomes systematically narrower in the smaller and larger radial bins. A feature of 
note in the 7.2 $<$ R $<$ 8.9 kpc bin, also seen in the corresponding H$_{2}$ map in Fig.~2, 
is the lack of emission in the range -140 $<$ l $<$ -100 deg. This gap may be a consequence of the 
structure of the Gould Belt and the corresponding Lindblad expanding ring. 
In the general direction of the gap, the Gould Belt is quite far from us, far enough that its gas would be shifted 
into the R=8.9-14 kpc bin (see, e.g., Fig.~5 in Perrot $\&$ Grenier 2003). In the highest radial bin 
(14 $<$ R $<$ 17 kpc) one sees clearly the well-known warping of the outer Galactic disk.

\subsection{Molecular hydrogen}

Templates of the radial distribution of H$_{\sc{2}}$ column densities have been 
constructed from CO data. In particular, we use the Galactic CO survey of Dame et al. (2001) 
except in a region covering parts of the first and fourth quadrants (
0$^{\circ}$ $<$ l $<$ 60$^{\circ}$ and 273$^{\circ}$ $<$ l $<$ 360$^{\circ}$, -5$^{\circ}$ $<$ b $<$ +4$^{\circ}$) 
where the NANTEN survey  (Fukui et al. 1999) is used. 
The NANTEN data set was obtained with the Nagoya University 4-m radio telescope 
located at Las Campanas in Chile. The final survey 
is a composite of 1,100,000 spectra at 4$^{'}$ spacing. The velocity resolution 
is 0.65 km/s although the version of the data we worked with has been 
slightly degraded to a resolution of 1 km/s. The velocity range is -250 
to +250 km/s. 
As for the Dame et al. (2001) data set, this has been obtained, although at different times, 
with two very similar
1.2-m telescopes, one in Cambridge, 
Massachusetts and the other at the Cerro Tololo Inter-American 
Observatory, Chile. The final composite CO data set has been 
constructed by interpolating all component surveys onto a uniform 
l-b-v cube with angular spacing of 0.125$^{\circ}$ and a velocity 
spacing of 1.3 km/s. In addition, 
bad channels and single missing spectra were filled in by interpolation. 
The new CO survey has 16 times more 
spectra than that of Dame et al. (1987), 3.4 times higher angular resolution and up to 10 
times higher sensitivity per unit solid angle. 
The proton column density in molecular hydrogen for a given ring $R_{i}$ and 
line of sight $k$ is then obtained (Lebrun et al. 1983) as:

\begin{equation}
(N^{i}_{H_{2}})_{k} = 2(2.8 \times 10^{20}) \int_{V^{i}_{LSR}} {T_{CO}(k,v)} dv \nonumber
\end{equation}

\noindent
Units are in 10$^{20}$ atoms cm$^{-2}$. 
In the expression above, following Giard et al. (1994), we assume a Galaxy-wide constant value of 
2.8 $\times$ 10$^{20}$ mol cm$^{-2}$ K$^{-1}$ km$^{-1}$ s for the ratio of H$_{2}$ column 
for the ratio of H$_{2}$ column density to velocity-integrated CO intensity (Bloemen et al. 1986). 
In addition, the integral is computed over the velocity range defined in eq.~(5). 

Fig.~2 shows the spatial maps of CO integrated over the same radial bins as the H$_{I}$ in Fig.~1. 
The most striking difference between the two  figures is the lack of CO emission in the 
highest radial bin, demonstrating the much larger radial extent of the Galactic H$_{I}$. Evident 
in the CO local bin (7.2 $<$ R $<$ 8.9 kpc) are well known molecular clouds such as the Cepheus flare 
($l$ $\sim$ 100 deg), the Aquila Rift ($l$ $\sim$ 20 deg) and the Chamaeleon ($l$ $\sim$ -60 deg). The 
Orion complex appears mainly in the 8.9 $<$ R $<$ 14 kpc bin.

\begin{figure*}
\centering
\includegraphics[width=18cm,height=16cm,angle=90]{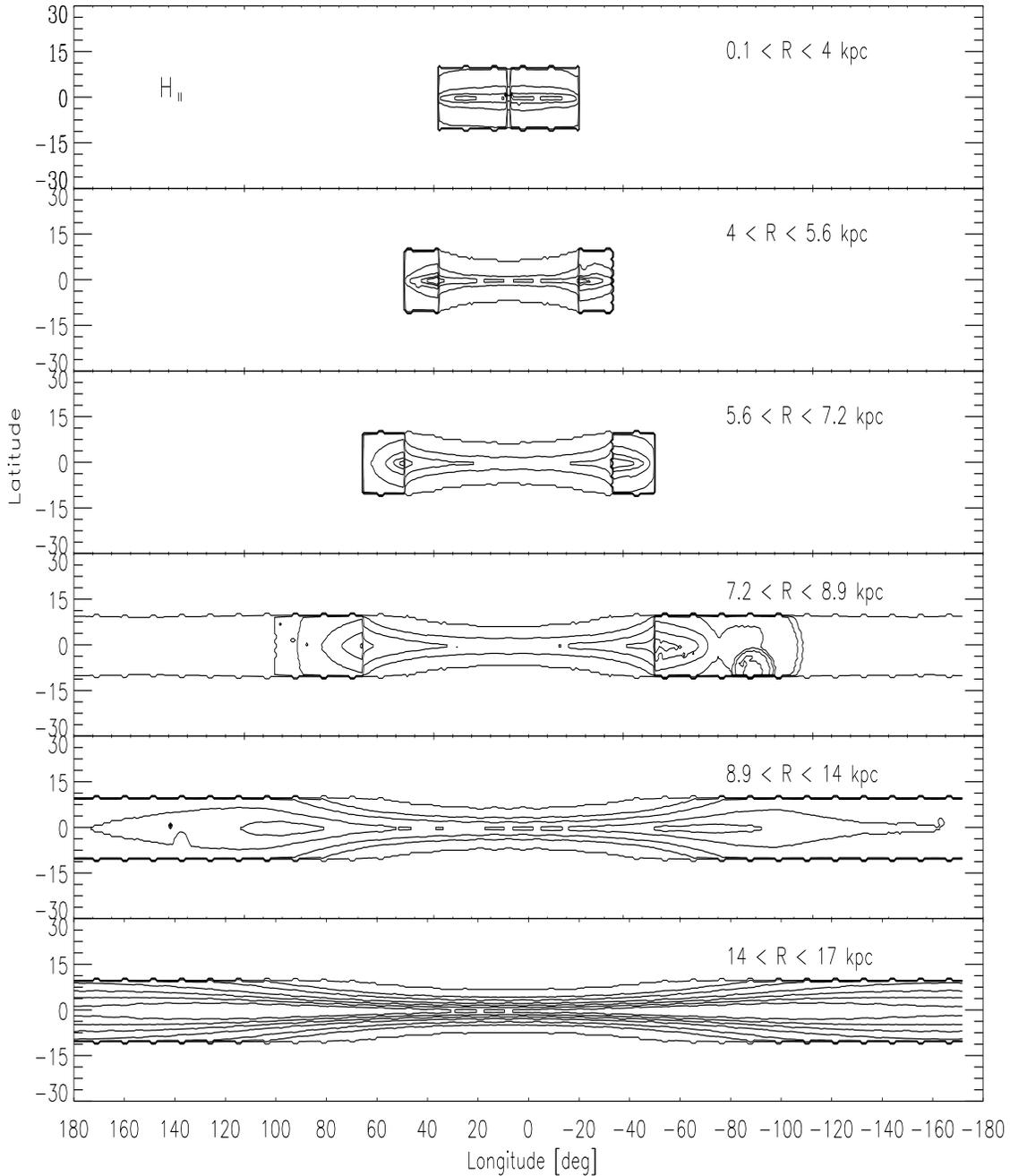}
\vspace*{0.3truecm}
\\
\caption{Ring 1 to 6 (from top to bottom) for H${_{\sc{II}}}^{diffuse}$ column density. Levels of the contours are: 
5,20,40,80,120,160,200,300 10$^{20}$ atoms cm$^{-2}$.}
\end{figure*}

\subsection{Ionized hydrogen}

\subsubsection{Previous works}
Before illustrating how the ionized gas has been handled in our analysis, 
we briefly review the approaches adopted by previous authors.   
Over the years, the contribution of ionized gas to the observed IR dust emission 
has been the subject of considerable debate. According to Mezger et al. (1982) and Cox et al. (1986) 60$\%$ to 
80$\%$ of the total infrared luminosity of the Galactic disk is contributed by dust associated 
with ionized gas. Sodroski et al. (1989) in their early analysis attempted to model 
this contribution by means of the 5 GHz radio continuum data of Haynes et al. (1978). To account 
for the non-thermal synchrotron component at this frequency, they estimated the thermal/non-thermal 
contributions at 1.4 GHz and then 
extrapolate that contribution to 5 GHz by adopted constant spectral indices, i.e. -0.1 for the thermal 
component and -0.7 for the non-thermal one\footnote{Quoted spectral indices assume S$_{\nu}$ $\propto$ $\nu^{\alpha}$.}. 
They concluded that less than 20$\%$ of the 5 GHz emission is of non-thermal origin. 
However, this method has a major problem. As shown by Paladini et al. 
(2005), a careful separation of the thermal and non-thermal components has to take into account the 
spatial variation of the synchrotron spectral index. When such variation is considered, 
one finds that the non-thermal emission contributes only $\sim$ 20$\%$ at $|b| \sim$ 
1$^{\circ}$. At higher latitudes, the synchrotron 
component increases significantly. In addition, it is shown that discrete H$_{\sc{II}}$ regions contribute 
about 10$\%$ of the total thermal emission. 
A slightly different approach is adopted by Sodroski et al. (1997). In their paper, the 5 GHz 
data are complemented with 19 GHz low-resolution (180$^{'}$) data from Boughn et al. (1991). 
As shown by Finkbeiner et al. (2004), the emission outside of H$_{\sc{II}}$ regions for $\nu \ge$ 14 GHz is 
likely dominated by spinning dust. Therefore the use of data at 19 GHz as a tracer of the ionized 
gas turns out to be rather problematic. 
On the contrary, Bloemen et al. (1990) exclude the ionized gas phase from their model 
based on the work by Cox and Mezger (1988), who concluded that dust associated with ionized gas 
contributes only 20$\%$ to the overall IR luminosity. As a consequence, bright H$_{\sc{II}}$ regions 
are masked on the basis of their 60$\mu$m/100$\mu$m colors and excluded from the fit. 
However, as noted by Giard et al. (1994), the 
large discrepancy between the IR observations and their best-fit model intensities is 
likely due to the exclusion of the ionized gas component from their analysis.

\subsubsection{This work}

\noindent
The best way to trace the ionized gas and its spatial distribution in the Milky Way 
would be a Galactic plane survey 
of radio recombination lines (RRLs). Since such a survey is not yet available, we have 
to resort to alternative methods. 
Taylor $\&$ Cordes (1993) have proposed a model which is based on
observations of pulsar dispersion and scattering measures (respectively, DM and SM).
The model has been incorporated and improved by Cordes $\&$ Lazio (2002, 2003, hereafter NE2001). 
The NE2001 model consists of five 
components: a large scale distribution, the Galactic Center,
the local ISM, individual clumps and individual voids. The large scale distribution is, in turn,
contributed by two axisymmetric components, i.e. a thin and a thick disk, and by
the spiral arms. For details about the functional form of these components, we
refer the reader to Cordes $\&$ Lazio (2002). By using the NE2001 model, we can construct 
rings (see Fig.~3) of electron column densities for the
diffuse ionized gas following the same guidelines as for the H$_{\sc{I}}$ and H$_{\sc{2}}$.
In order to check the reliability of this model, it would be highly desirable to tie up the
computed column densities with observations. This corresponds in practice to
converting column densities (DM) into emission measures (EM). Although some information exists 
on the relation between these two quantities (see, for example, Mitra, Berkhuijsen, Muller (2003)), 
it is still too uncertain to allow a realistic conversion and therefore to validate 
and confidently use the model. \\

Another way to trace the ionized gas is through its free-free emission. With the 
release of WMAP data, all-sky templates
of the Galactic foregrounds have become available (Bennett et al., 2003b). These include
synchrotron, free-free and dust emission maps. Such maps have been obtained with
the Maximum Entropy Method (MEM). With this technique, prior templates for each component
of Galactic emission are considered and it is assumed that the observed signal at each frequency is 
contributed by four terms, namely, CMB, synchrotron, free-free and dust. For free-free
emission (i.e. thermal bremstrahlung emission produced by ionized gas), the H$\alpha$ map of 
Finkbeiner (2003) is used as a template. We note that H$\alpha$ maps cannot be used
directly to trace the ionized gas given that they suffer heavily from obscuration due to dust
absorption at low latitudes ($|b| <$ 5 deg).  
As reported in Bennett et al. 2003b, all
the uncertanties associated with the H$\alpha$ map (i.e., errors due to correction for dust as well as to
conversion from H$\alpha$ to free-free, etc.,) combine together and do not allow one to put
tight constraints on the recovered free-free map. Despite these drawbacks and the fact 
that the WMAP free-free map traces continuum emission and, as such, does not allow the recovery of  
the spatial location of the emitting gas, we have decided to make use of it for our analysis, 
convinced that it represents the best available option. The WMAP team has released  
free-free maps at each WMAP band. We have used the one in the Ka band (i.e. 33 GHz) as a  
compromise between spectral dependence and angular resolution. Such a map, provided in  
units of thermodynamic mK, has been converted into units of 10$^{20}$ electrons cm$^{-2}$ by 
applying eq.~(8) in Sodroski et al. (1989) with a value of 10 cm$^{-3}$ assumed for $n_{eff}$.
We have compared the performance of the WMAP free-free map with respect to the NE2001 model.  
The result, in terms of reduced $\chi^{2}$, i.e. $\overline{\chi}^{2}$, for a sample of wavelengths is shown in Table~1. 

\begin{table}[h]
\begin{center}
\begin{tabular}{ccccc}
\hline
\hline
    & 60 $\mu$m & 100 $\mu$m & 140 $\mu$m & 240 $\mu$m  \\
\hline
\\
WMAP free-free  &    1.087   &  1.046 & 1.037 & 1.033 \\
NE2001          &    1.141   &  1.078 & 1.056 & 1.043 \\
\\
\hline
\hline
\end{tabular}
\caption{$\overline{\chi}^{2}$ comparison for WMAP free-free map and NE2001 electron density model. 
}
\end{center}
\end{table}

\noindent
Clearly, the fit to the observed emission gets worse when the NE2001 model is applied. This is expected 
since the NE2001 model mostly accounts for the 
very large-scale structure while, as noted by Sodroski et al. (1989), the bulk of the 
IR emission consists of a distribution of prominent, discrete sources superposed on a 
continuous background.\\

The WMAP free-free map is characterized by a rather low angular resolution (1$^{\circ}$). This 
implies that most of the sources (H$_{\sc{II}}$ regions)  which contribute to the Galactic thermal emission 
are highly diluted since their average angular size is $\sim$ 6$^{'}$. 
Although the diffuse component accounts for nearly all of the ionized gas in the Galactic disk 
(its mass is a factor of 20 larger than that of classical H$_{\sc{II}}$ 
regions (Reynolds 1991)), we have included the contribution of discrete sources for completeness. For this purpose, 
we have made use of the
Paladini et al. (2003) catalog (hereafter Paper I).  This catalog contains 1442 Galactic H$_{\sc{II}}$ regions
for which angular diameter and flux densities at a reference frequency (2.7 GHz) are given.
Radio recombination lines have been observed for $\sim$ 800 of these. This kinematic information,
combined with the Fich, Blitz $\&$ Stark (1989) rotation curve, has been used in Paladini, Davies $\&$
DeZotti (2004) (hereafter Paper II) to compute galactocentric and solar distances. Only sources (575)
for which the observed velocity is $|V_{obs}| \ge$ 10 km/s have been considered, to minimize
uncertainties in the derived distances due to peculiar motions. For 281 H$_{\sc{II}}$ regions located inside the
solar circle, the distance ambiguity has been resolved by applying one of these methods:
\begin{enumerate} \item{comparison with auxiliary (optical or absortion) data when these are
available} \item{use of a statistically significant luminosity-physical diameter
correlation\footnote{See eq.~(6) in Paper II}} \end{enumerate}

\noindent
Following these guidelines, we obtain 550 H$_{\sc{II}}$ regions for which the solar distance is
uniquely determined. In Paper I we have also investigated the completeness of the data set. 
Such a study is complicated by the fact that the catalog is constructed by combining data 
obtained
with different instruments, observing at different frequencies and angular resolution.
Based on simple geometric arguments, we estimate that the catalog is
complete, at 2.7 GHz,
for flux densities $\geq$ 7 Jy (see Fig.~1 of Paper I). For fainter sources, the catalog completeness is 
mostly limited by confusion. For each source, we derive the electron density 
according to (Schraml $\&$ Mezger, 1969): 

\begin{equation}
n_{e} = 98.152\times \nu^{0.05} \hskip 0.1 truecm T_{e}^{0.175} \hskip 0.1 truecm S^{0.5} 
\hskip 0.1 truecm D^{-0.5} \hskip 0.1 truecm \theta^{-1.5} \hskip 0.5truecm [cm^{-3}]
\end{equation}

\noindent
being $\nu$ the frequency in GHz, $T_{e}$ the electron temperature in K, $S$ the flux density in 
Jy, $D$ the solar distance in kpc and $\theta$ the angular size in arcmin.  
From $n_{e}$ we then compute the electron column density simply by multiplying for 
the linear diameter expressed in pc. 
For most of the sources, an electron temperature has been computed from the observed 
recombination
line. However, when $T_{e}$ is not available, we estimate it by
applying the empirical relation: 

\begin{equation}
T_{e} (R) = (4166 \pm 124) + (314 \pm 20) R \hskip 1.5truecm [K]
\end{equation}

\noindent
as derived in Paper II, with $R$, the galactocentric radius, expressed in kpc. As can 
be seen in Fig.~4 of Paper II, the 
$T_{e}-R$
correlation is affected by a large scatter. However, eq.~(2) has been used only for $\sim$ 20$\%$ 
of the sources.

\begin{figure*}
\includegraphics[width=8.8cm,height=10.85cm]{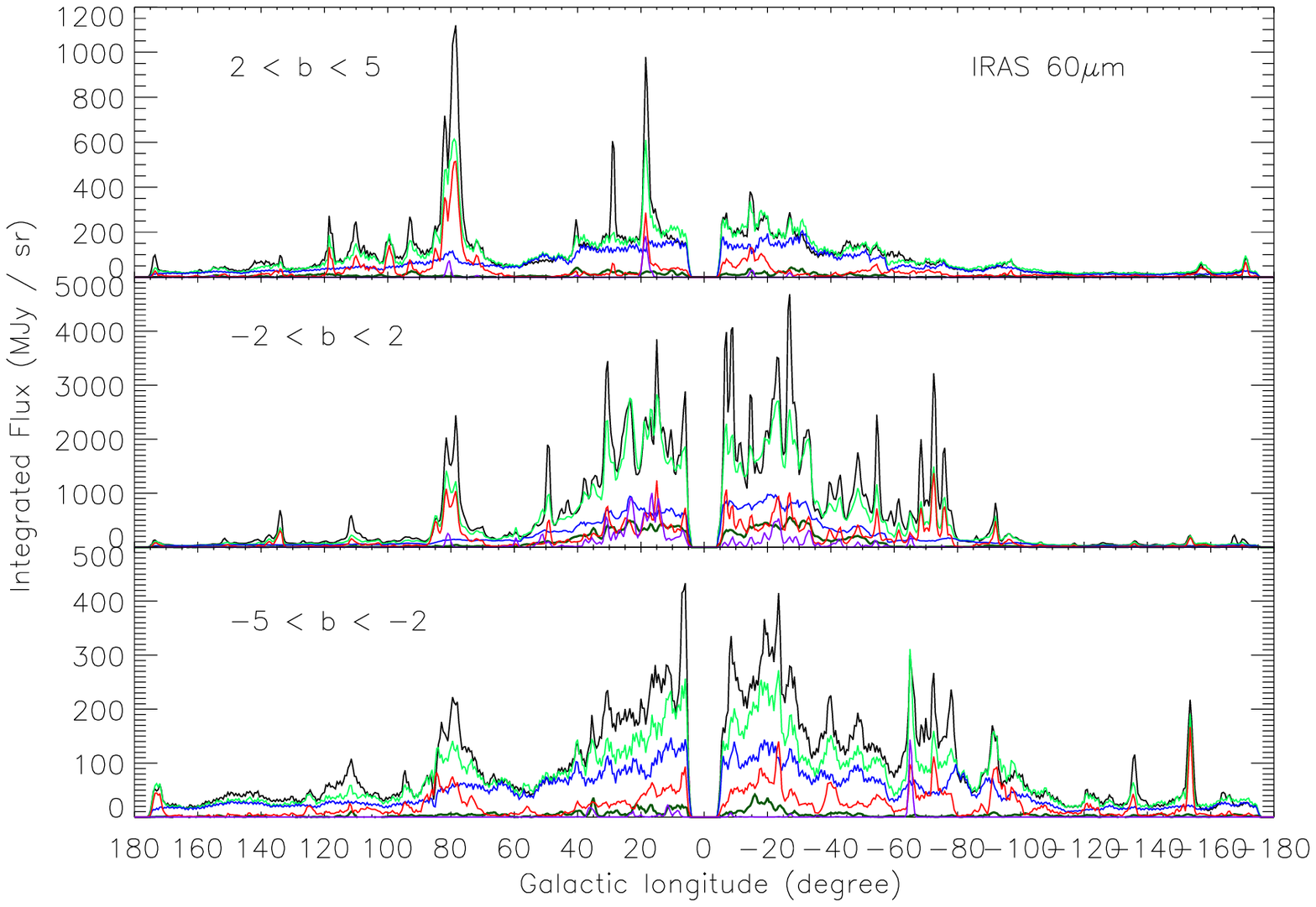}
\includegraphics[width=8.8cm,height=10.85cm]{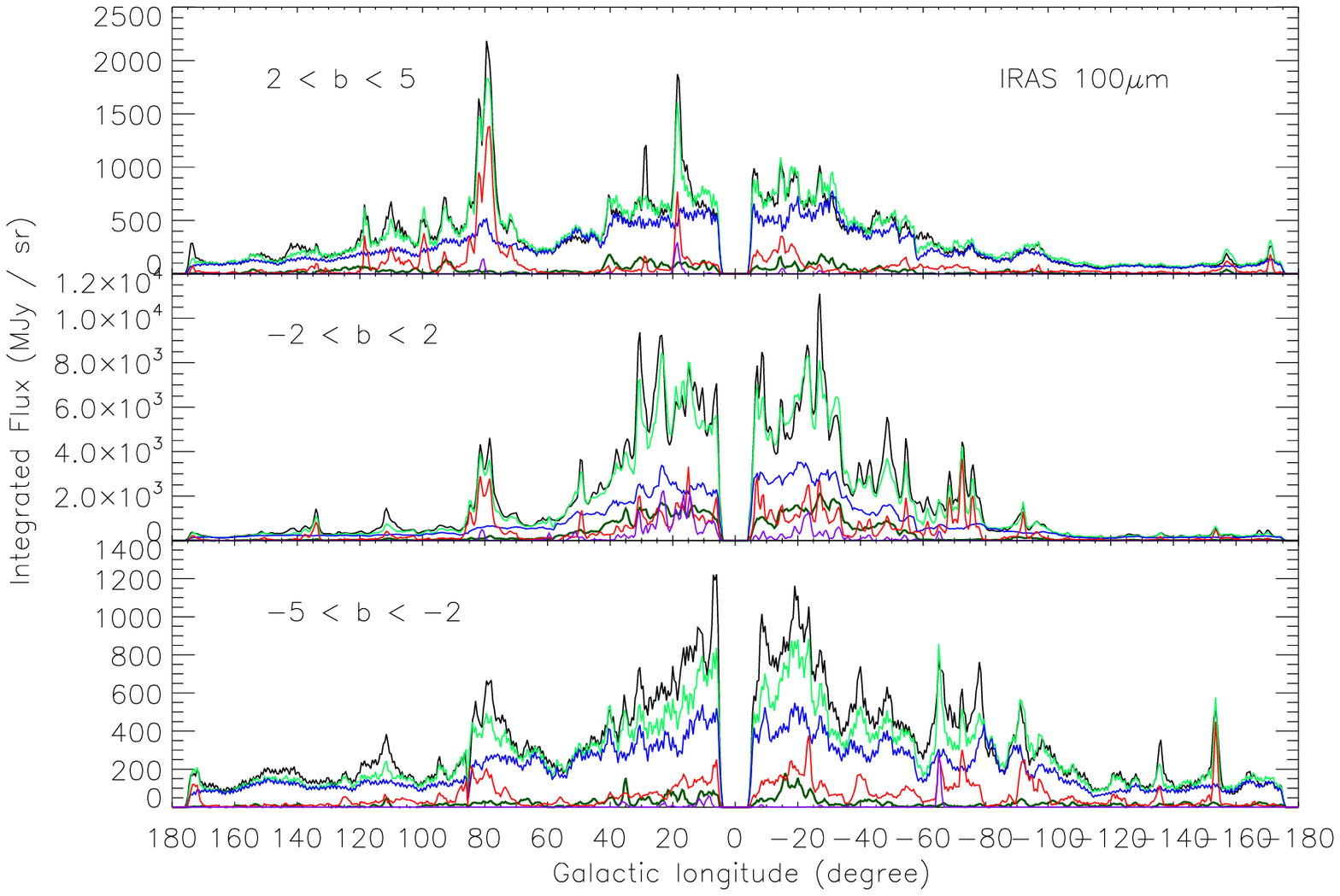}\\
\includegraphics[width=8.8cm,height=10.85cm]{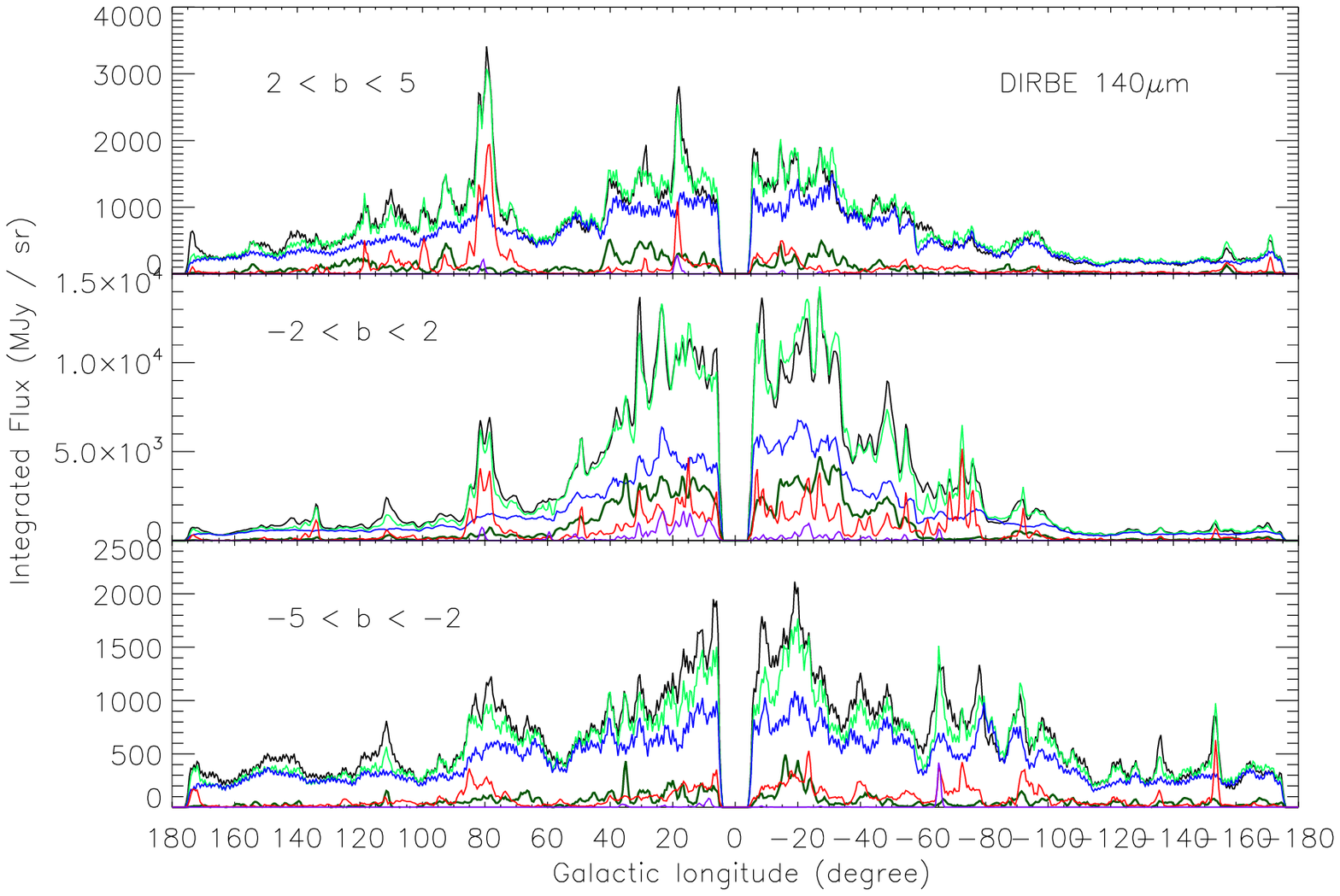}
\includegraphics[width=8.8cm,height=10.8cm]{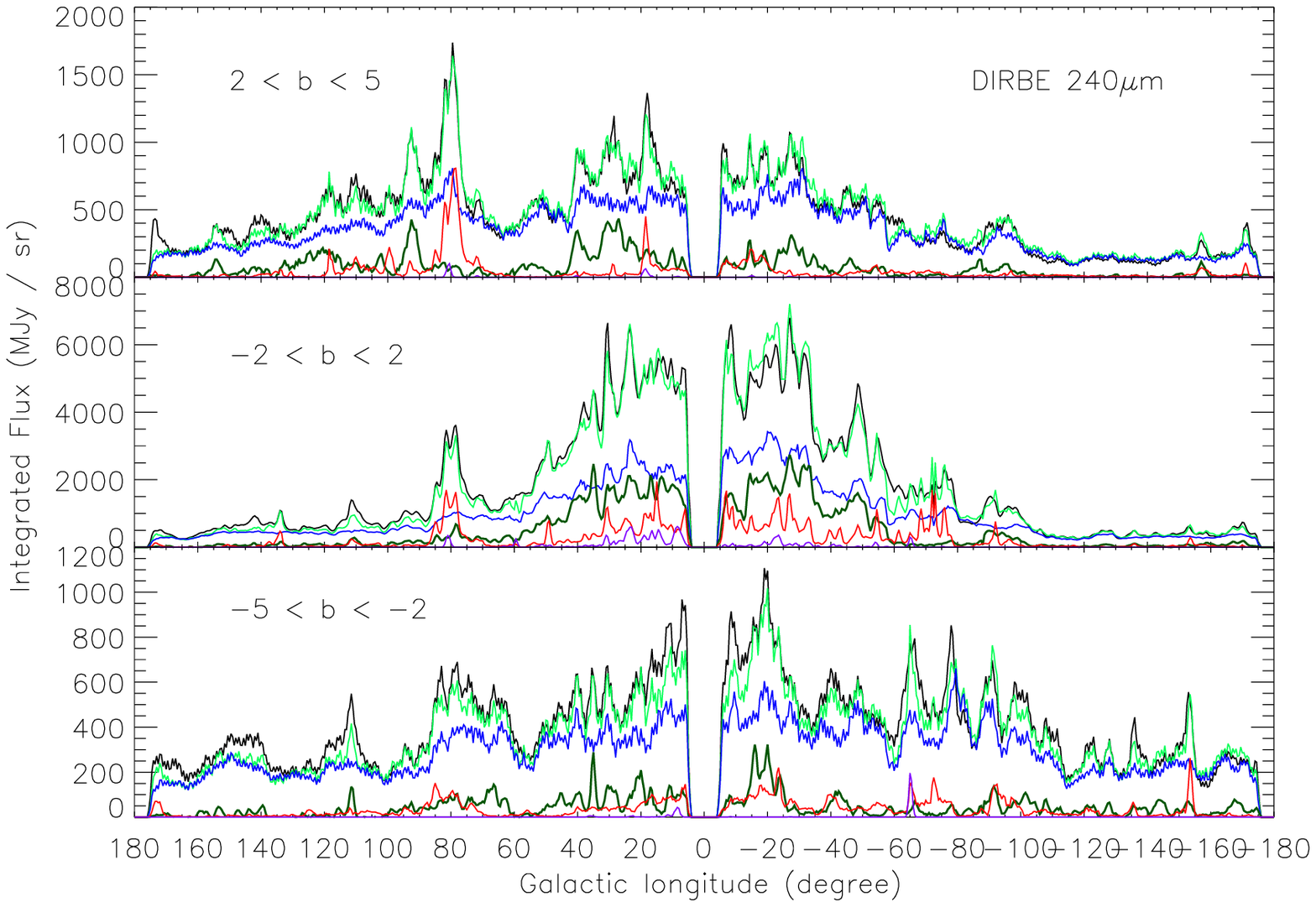}\\
      \caption{Longitude profiles for IRAS 60 and 100 $\mu$m (top panels) and DIRBE 140 and 240
$\mu$m (bottom panels). Black line denotes observed emission while light green line corresponds to the
fitted model. In addition, blue line is for H$_{\sc{I}}$, dark green line for H$_{\sc{2}}$, red line for H${_{\sc{II}}}^{diffuse}$ and
magenta line for H${_{\sc{II}}}^{compact}$. Profiles are obtained by averaging over the shown
latitude intervals. The apparent noise is not instrumental: along the Galactic plane, the S/N is very high 
and error bars representing the instrumental noise would be of the same size of the plotting line.}
\end{figure*}

\begin{figure*}
\includegraphics[width=8.8cm,height=10.8cm]{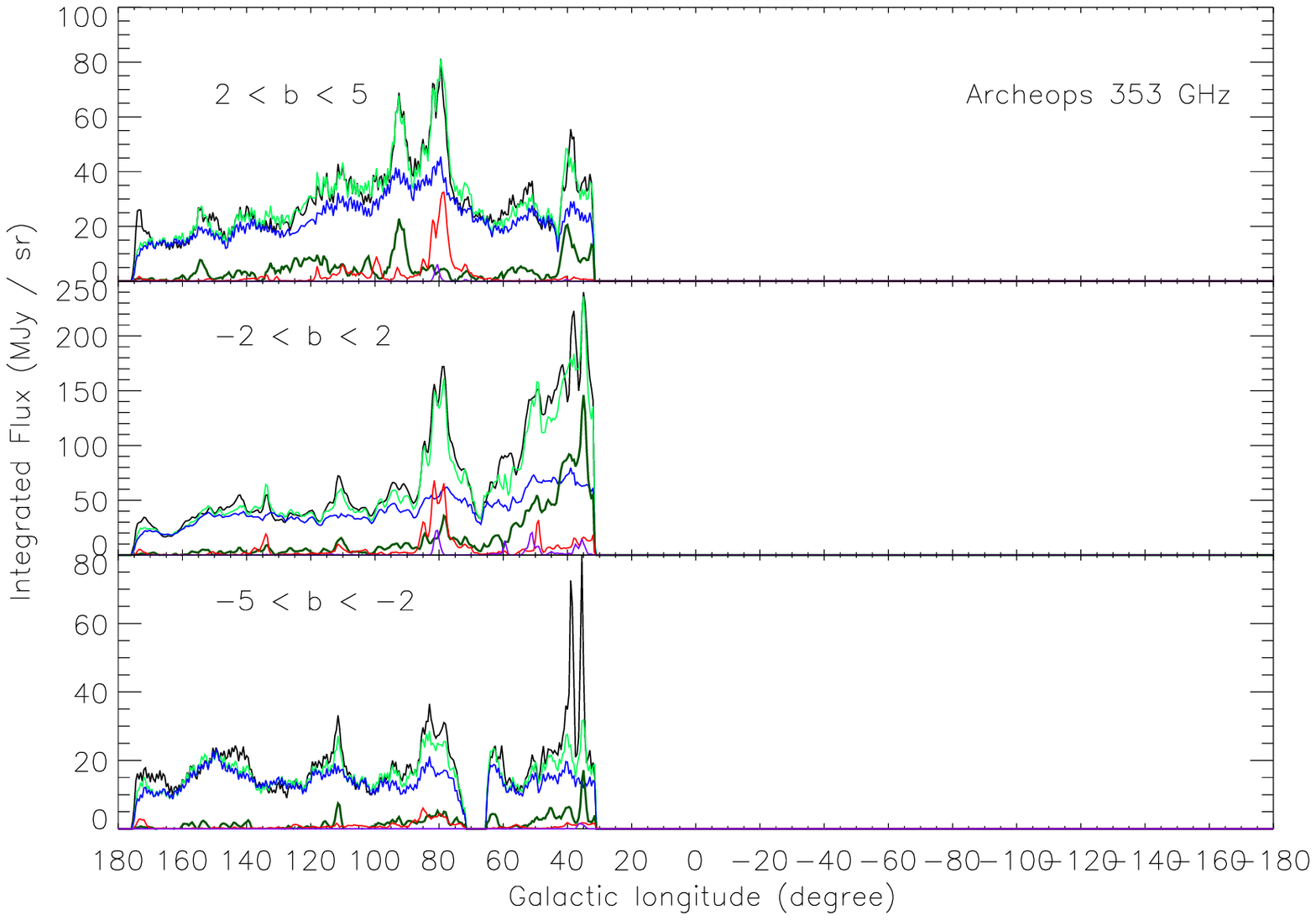}
\includegraphics[width=8.8cm,height=10.8cm]{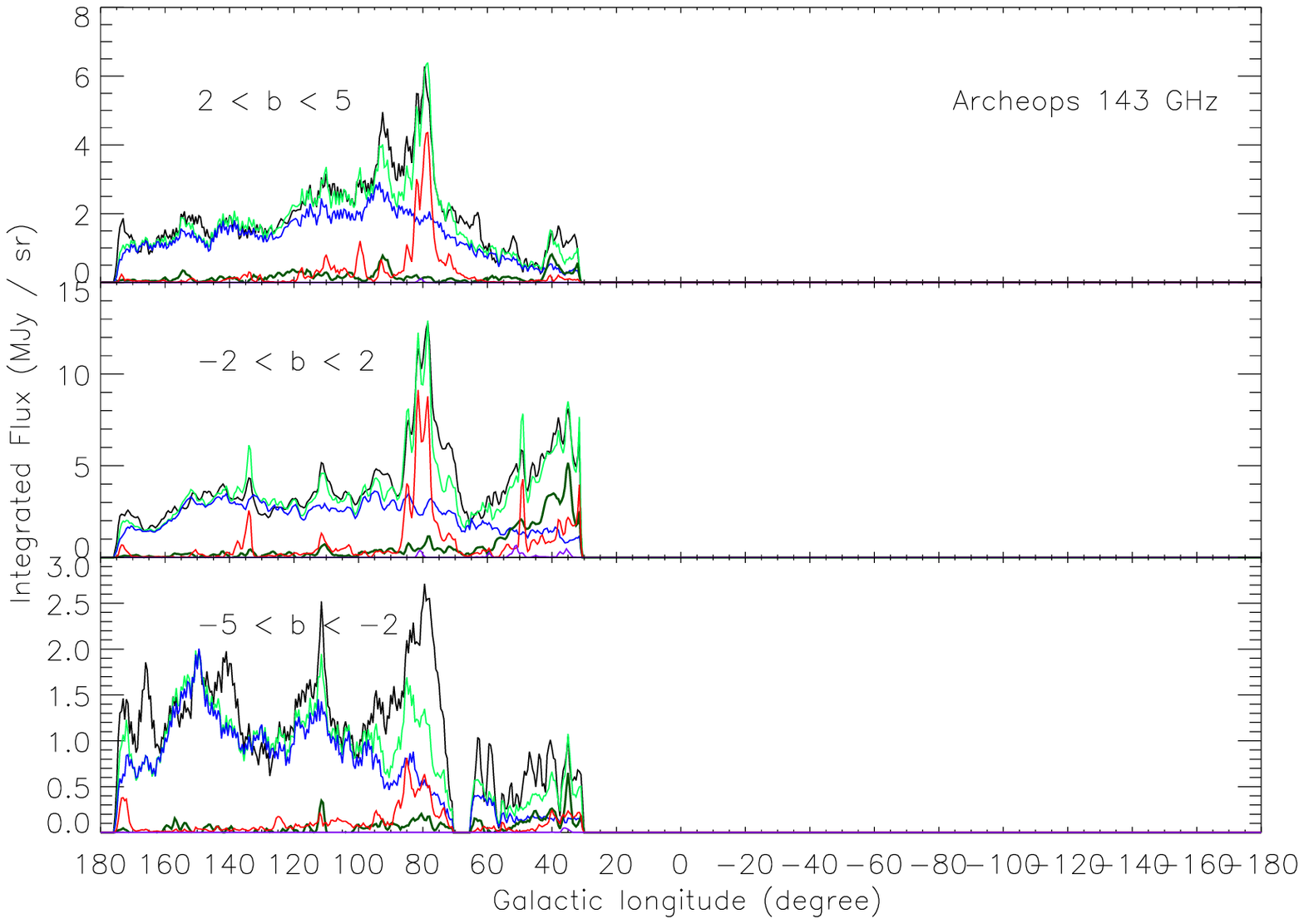}\\
\includegraphics[width=8.8cm,height=10.8cm]{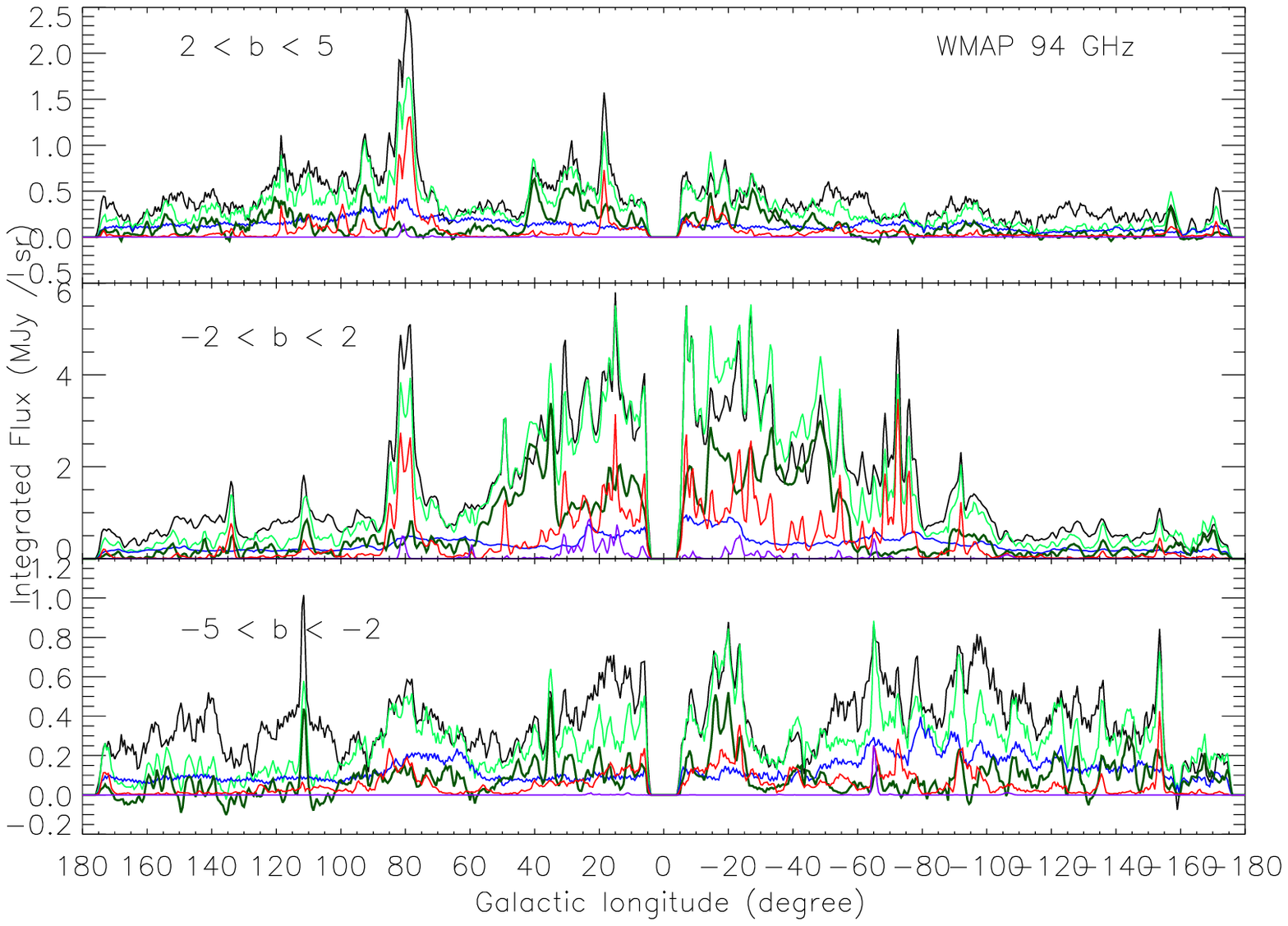}
\includegraphics[width=8.8cm,height=10.8cm]{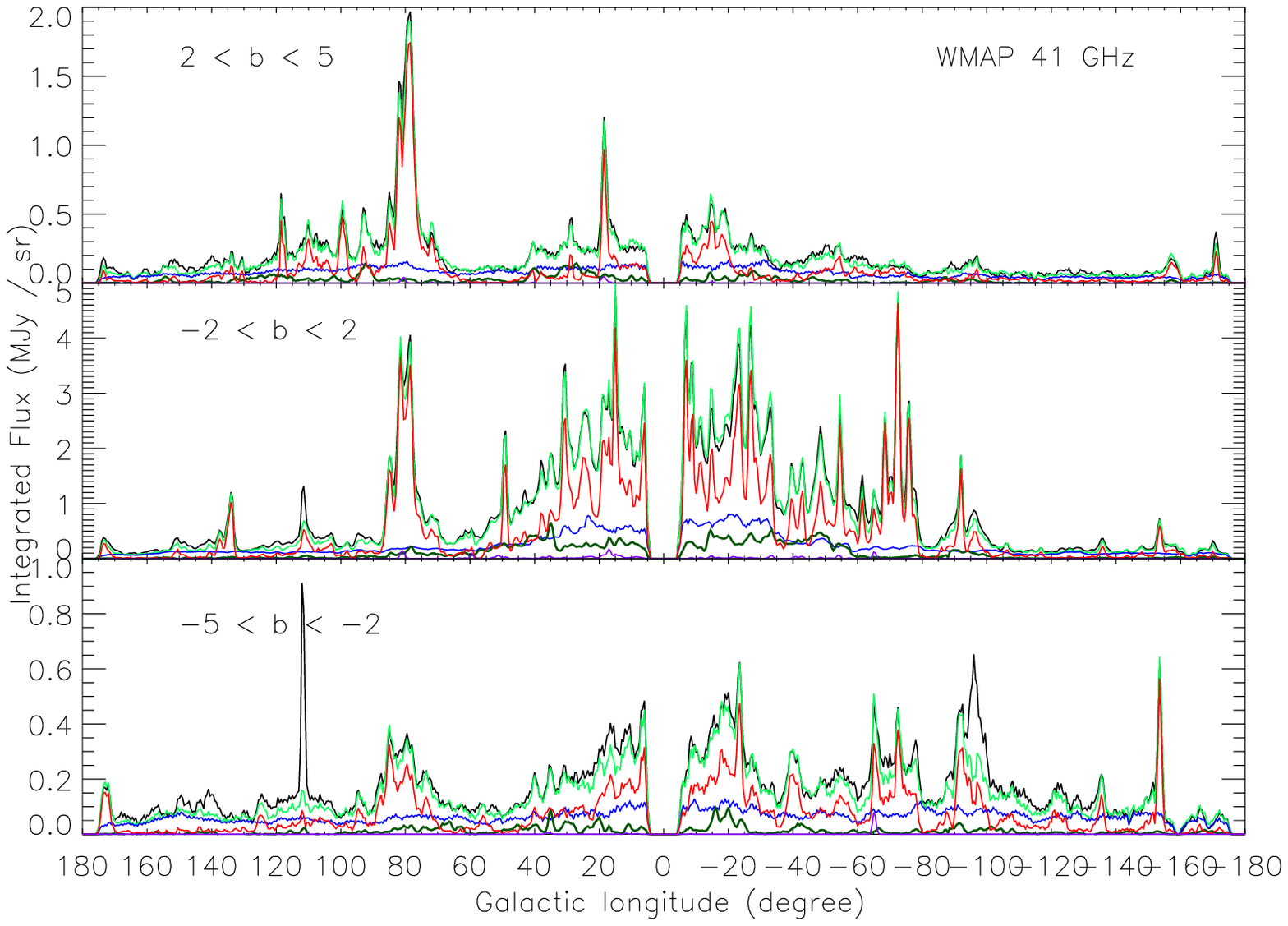}\\
      \caption{Longitude profiles for Archeops 850 and 2096 $\mu$m (top panels) and WMAP W and Q bands
(bottom panels). Archeops data base covers only the longitude interval 30$^{\circ}$ $< l <$ 180$^{\circ}$.
Black line denotes observed emission while light green line corresponds to the
fitted model. In addition, blue line is for H$_{\sc{I}}$, dark green line for H$_{\sc{2}}$, red line for H${_{\sc{II}}}^{diffuse}$ and
magenta line
for H${_{\sc{II}}}^{compact}$. Profiles are obtained by averaging over the shown
latitude intervals. The same remark as in Fig.~4 on instrumental noise applies here.}
\end{figure*}

\section{Derived physical parameters}

\subsection{The iterative inversion method}

In matrix form, the minimization of eq.~(2) is equivalent to solving the linear system:

\begin{equation}
(A^{t}\Pi A) \epsilon_{\lambda} = A^{t}\Pi I_{\lambda}
\end{equation}

\noindent
where A is a $(3n+1)\times N$ matrix of the type:

\hskip -0.3truecm
\[ \left( \begin{array}{cccccccccc}
N^{1,1}_{H_{I}}&..&N^{1,n}_{H_{I}} & 2N^{1,1}_{H_{2}}&..&2N^{1,n}_{H_{2}} & N^{1,1,c}_{H{_{II}}}&..&N^{1,n,c}_{H{_{II}}} & N^{1,d}_{H{_{II}}}\\
:          &     &     :      &     :      &     &    :   &     :      &     &    : &  :   \\
N^{N,1}_{H_{I}}&..&N^{N,n}_{H_{I}} & 2N^{N,1}_{H_{2}}&..&2N^{N,n}_{H_{2}} & N^{N,1,c}_{H{_{II}}}&..&N^{N,n,c}_{H{_{II}}} & N^{N,d}_{H{_{II}}} \end{array} \right)\] 

\noindent
$n$ being the number of rings in the decomposition and $N$ the number of pixels. In the expression above, $N^{j,i,c}_{H{_{II}}}$ 
denotes the column density in pixel $j$ and ring $i$ for the source (or {\em{compact}}) component of the ionized gas, while $N^{j,d}_{H{_{II}}}$ 
is the corresponding column density for the {\em{diffuse}} component. $\Pi$ has dimension $N \times N$ and is given by:

\[ \Pi = \left( \begin{array}{cccc}
\frac{1}{{\sigma_{1}}^{2}} &  0 &  ... & 0 \\
0    &        &                &    : \\
:    &        &                &  0 \\
0    &  ...      &        0        & \frac{1}{{\sigma_{N}}^{2}}\end{array} \right)\]

\noindent 
with $\sigma_{j}$ the pixel noise of the input map. In eq.~(11), the correlation matrix 
$A^{t}\Pi A$ is symmetric and positive-defined which allows the linear system to be solved 
by means of a Cholesky inversion method (see, for instance, Golub $\&$ van Loan, 1989). The major difficulty is represented by a correct 
evaluation of the noise which is largely dominated by the uncertainty associated 
with the model. To circumvent such a problem, we have applied an iterative method. This 
consists of making an initial assumption about the (instrumental) noise, computing the corresponding solution 
and then using the noise derived from this partial solution to 
reiterate the process.   
A true estimate of the real noise is obtained when, by repeating this process several times, the system converges. 
Convergence is, on average, reached within 10 iterations.  
In the first iteration the (intrumental) noise per pixel is assumed to be given by:

\begin{equation}
\sigma^{1}_{j}=\sigma_{ini}+5\% I_{\lambda}(j)
\end{equation}

\noindent
where $\sigma_{ini}$ represents the average noise of the I$_{\lambda}$ map at high latitudes and 
I$_{\lambda}$(j) is the value of the map in pixel $j$. In this expression, the term proportional to 
the observed signal is very important. The sky is in fact highly inhomogeneous and variations in signal 
in the range 1 to 10,000 (in brightness units) are observed when high latitutes are compared to low ones. It is then very unlikely 
that the difference between the true and the measured signal is only the same additive noise as it is at high Galactic 
latitudes. A proportional term taking into account the long term drift of the instrument gain needs to 
be considered. Nonetheless, it is important to emphasize that such definition 
of the noise is still rather crude and purely operational given the iterative method which has been adopted.  
\\ 
Emissivities coefficients obtained by following the procedure now described are shown 
in Table~2.

\begin{table*}[h]
\begin{center}
\begin{tabular}{cccccccc}
\hline
\hline
\\
{\em Gas phase} &  Ring  &  \multicolumn{6}{c}{$\epsilon{_\nu}$ \hspace{0.3cm} [MJy sr$^{-1}$ (10$^{20}$ H atoms cm$^{-2}$)$^{-1}$]} \\
                &  (kpc)      &   60 $\mu$m & 100 $\mu$m & 140 $\mu$m & 240 $\mu$m & 850 $\mu$m & 2096 
$\mu$m \\
\hline
\\
H$_{I}$         &  0.1-4      &  3.9$\pm$0.08  &  13.7$\pm$0.89   &    26.4$\pm$0.68    &   
12.46$\pm$0.27    &   -----    &   ----     \\
                &  4-5.6      &   1.53$\pm$0.07  &      5.37$\pm$0.19   &   9.54$\pm$0.23  &  
4.52$\pm$0.087    &   0.087$\pm$0.050   &  ----      \\
                &  5.6-7.2  & 0.57$\pm$0.01  &  2.26$\pm$0.04  &  4.51$\pm$0.08 & 
2.54$\pm$0.04 &  
0.10$\pm$0.004  & ----       \\
                &  7.2-8.9  &  0.32$\pm$0.001  &  1.37$\pm$0.01   & 3.02$\pm$0.01 & 
1.95$\pm$0.02  & 0.09$\pm$0.002  & 0.001$\pm$0.0002       \\
                &  8.9-14  &   0.061$\pm$0.001  & 0.39$\pm$0.003  & 1.06$\pm$0.014  &  
0.90$\pm$0.0046  & 0.091$\pm$0.0006  & 0.01$\pm$4.8e-05 \\
                &  14-17    &   0.01$\pm$0.004 &   0.07$\pm$0.01    &   0.38$\pm$0.05  & 
0.45$\pm$0.03   & 0.04$\pm$0.002   &   ----     \\
\\
H$_{2}$        &  0.1-4 & 0.12$\pm$0.04  &   0.17$\pm$0.11  &  0.23$\pm$0.12   & 0.18$\pm$0.09            
& ---- &  ----  \\
                & 4-5.6  &  0.20$\pm$0.01 &  0.74$\pm$0.04  &  1.45$\pm$0.05  & 0.75$\pm$0.04           
&   0.05$\pm$0.007  & 0.001$\pm$0.0004       \\
                &   5.6-7.2  & 0.14$\pm$0.01  & 0.63$\pm$0.02  &  1.72$\pm$0.03  & 
1.10$\pm$0.04  
&  0.051$\pm$0.006  & 0.00014$\pm$0.0003     \\
                &  7.2-8.9 & 0.04$\pm$0.002  & 0.18$\pm$0.005  &  0.64$\pm$0.005  & 
0.58$\pm$0.010   &   0.031$\pm$0.0005 &  0.001$\pm$8.4e-05      \\
                &  8.9-14  &  0.03$\pm$0.001  &  0.15$\pm$0.004   &  0.47$\pm$0.02  & 
0.42$\pm$0.014  & 0.02$\pm$0.001 & 0.001$\pm$0.0001      \\
                &   14-17 &  0.009$\pm$0.004  &   ----  &   ----  & 0.0052$\pm$0.02  &  
0.017$\pm$0.002 &  0.0026$\pm$0.0002      \\
\\
H$_{II}^{compact}$      &   0.1-4  &  0.18$\pm$0.02  &  0.88$\pm$0.14  &  1.23$\pm$0.26   & 0.64$\pm$0.14 
&  
----  
& ---- \\
                &  4-5.6 &  2.44$\pm$0.24   & 5.93$\pm$0.29  &  4.22$\pm$0.43   &  1.52$\pm$0.32  
&  ---- 
&  ----  \\
                &  5.6-7.2    &  1.14$\pm$0.11  &  1.83$\pm$0.22 & 1.75$\pm$0.32  & 0.38$\pm$0.12 
&  
0.097$\pm$0.02 &    0.003$\pm$0.001 \\
                &    7.2-8.9   & 3.49$\pm$0.31  &  6.81$\pm$0.68  &   10.17$\pm$1.45 & 
4.77$\pm$1.1  & 
0.26$\pm$0.06  & 0.005$\pm$0.001       
\\
                &  8.9-14   &  ----  &    ----    &  ----    &  ----   & 0.024$\pm$0.02    & 
----    
\\
                &    14-17  & 0.29$\pm$0.04   &  0.071$\pm$0.15  & ----  & ----  &   ----  &  
----\\
\\
H$_{II}^{diffuse}$              &   0.1-17  &   9.3$\pm$0.16  &  24.9$\pm$0.32  &   35.1$\pm$0.64  &  
14.6$\pm$0.38     
&   0.58$\pm$0.007   & 0.07$\pm$0.002   \\
\\
\hline
\hline
\end{tabular}
\caption{ Conversion factors to 4$\pi\nu\epsilon_{\nu}$ are, in units of 10$^{-31}$ W (H 
atom)$^{-1}$: 6.28, 3.77, 2.68, 1.57, 0.44, 0.18 at 60, 100, 140, 240, 850 and 2096 $\mu$m 
respectively. Quoted errors are derived as described in Section 4.}
\end{center}
\end{table*}

\begin{figure*}[h]
\includegraphics[width=7cm,height=5.7cm]{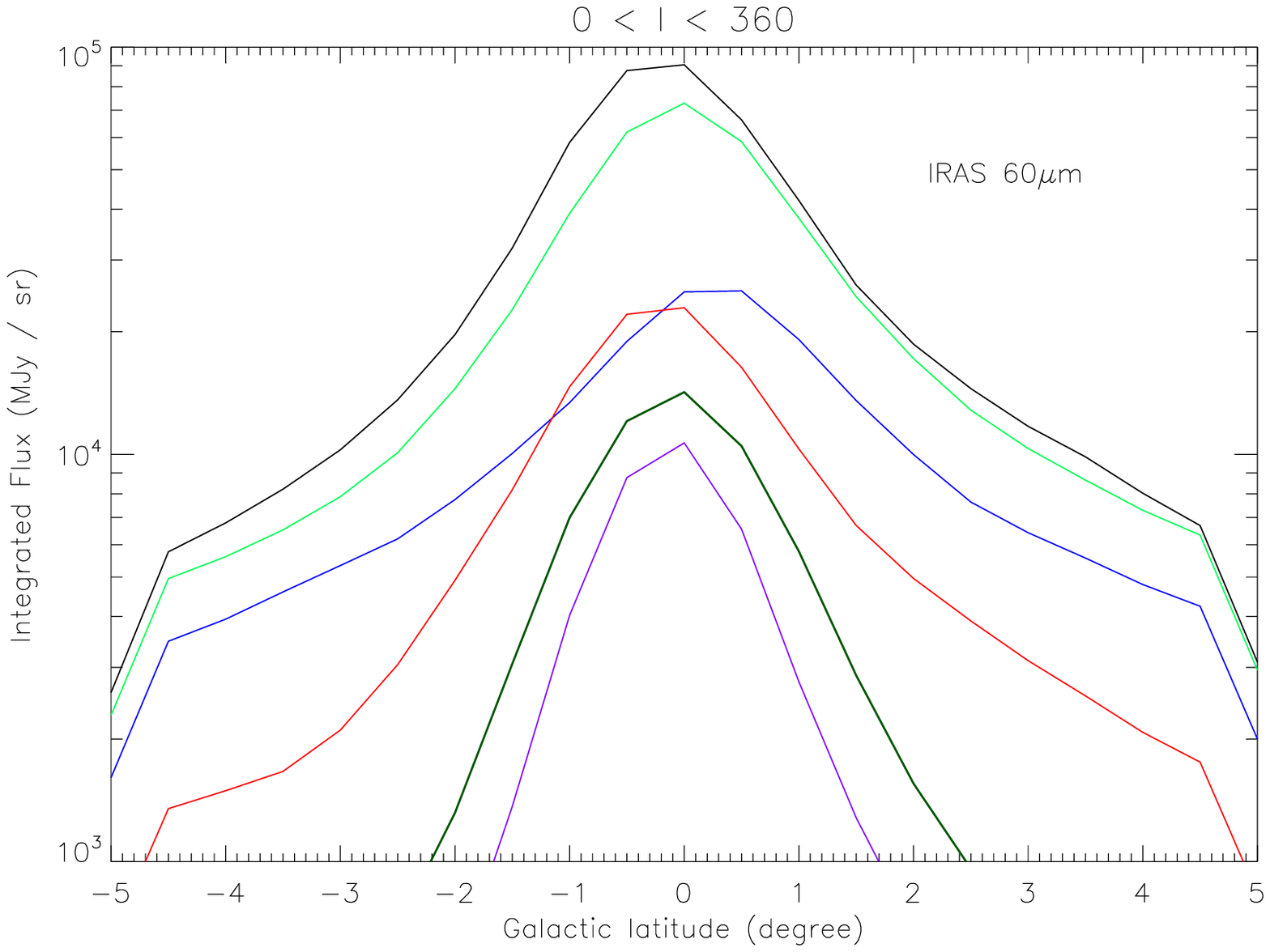}
\includegraphics[width=7cm,height=5.7cm]{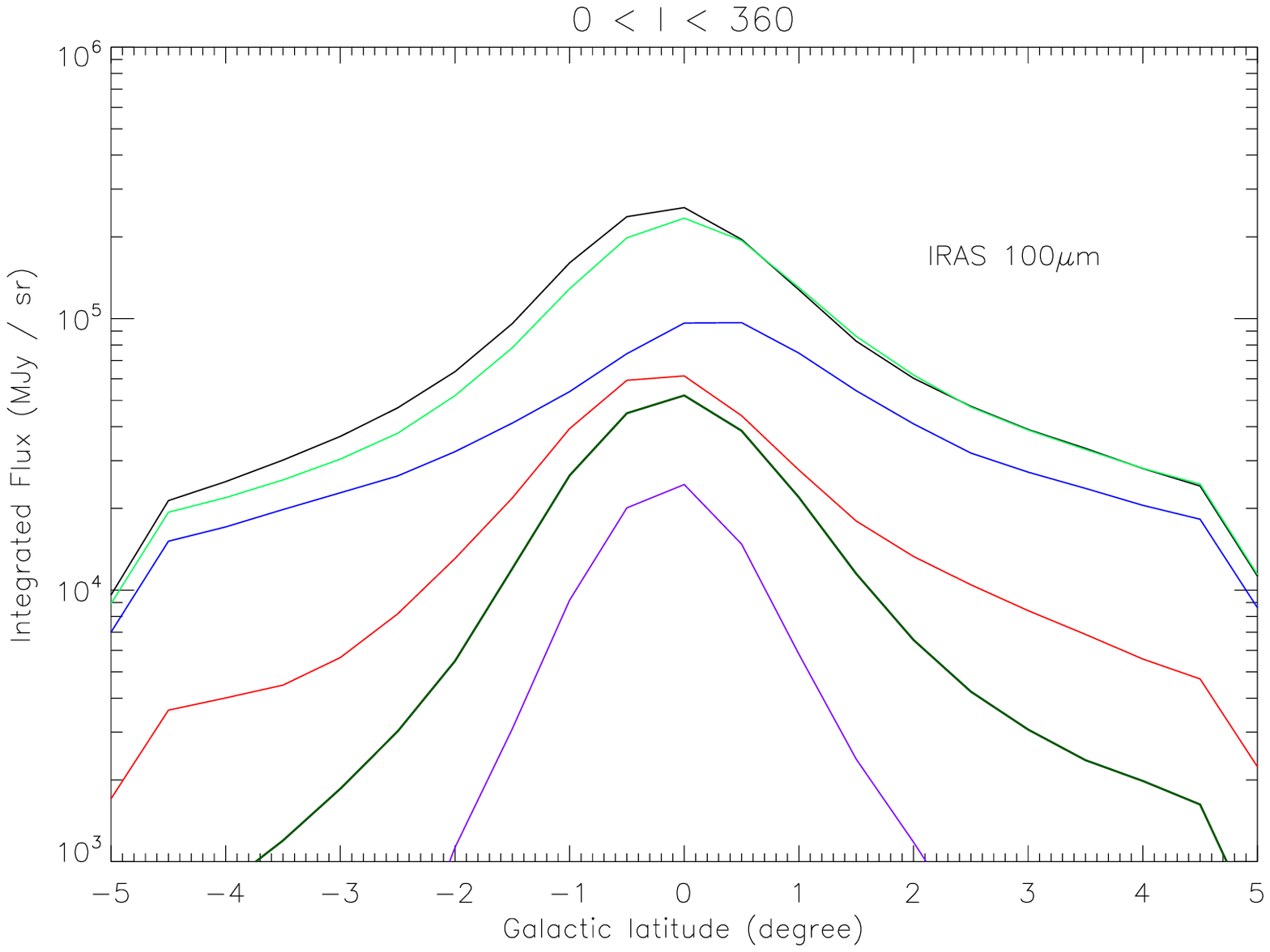}\\
\includegraphics[width=7cm,height=5.7cm]{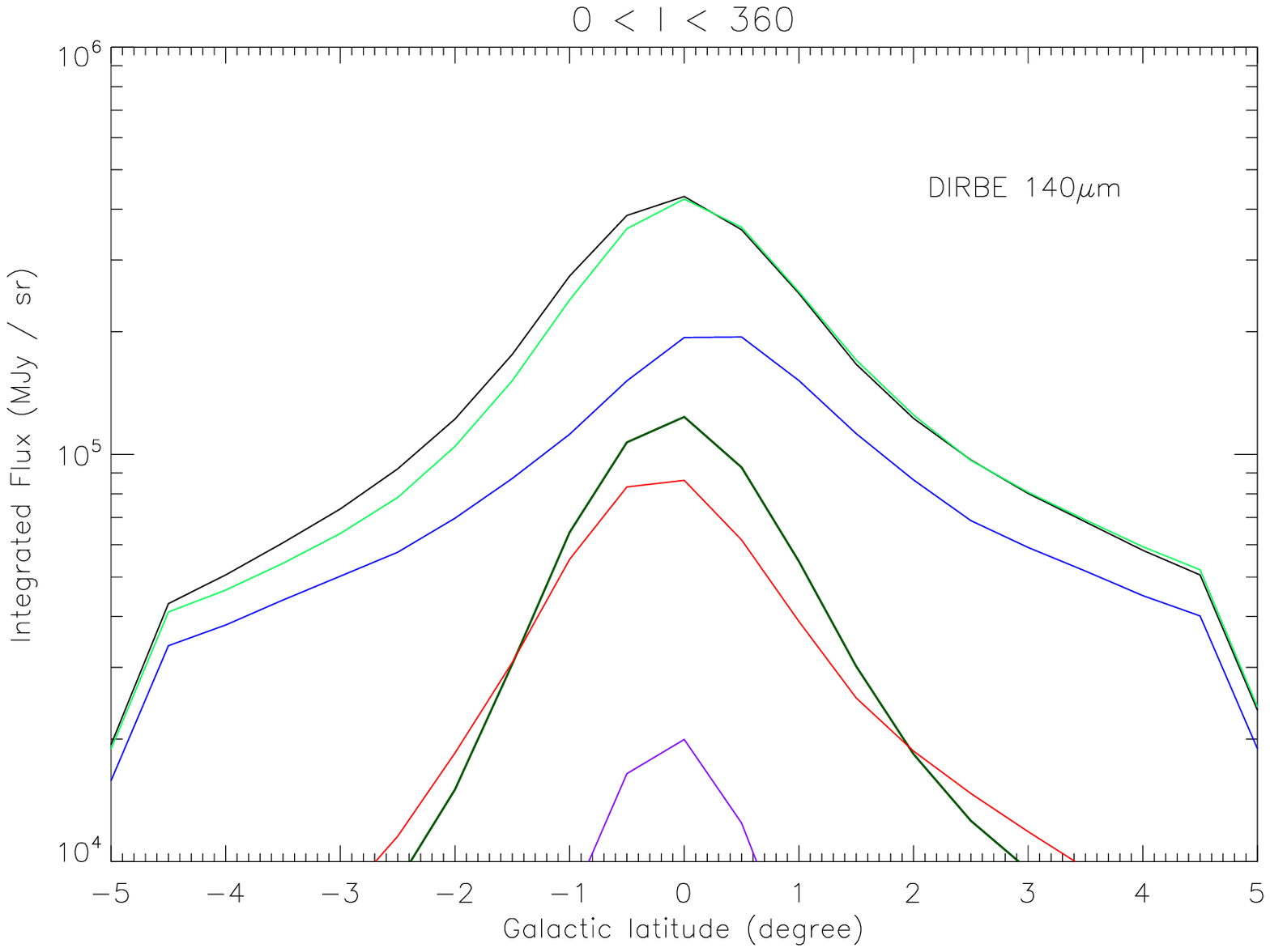}
\includegraphics[width=7cm,height=5.7cm]{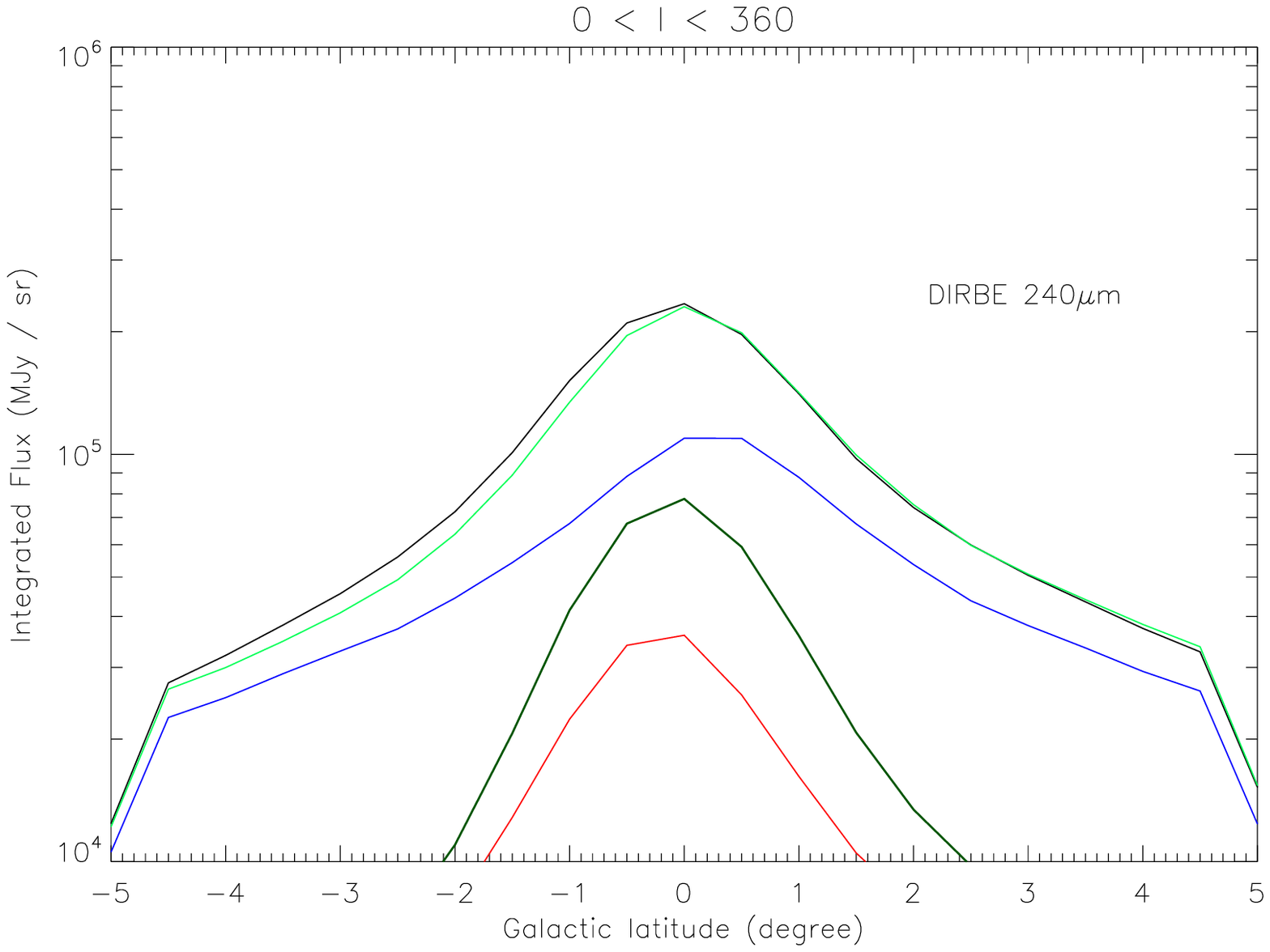}\\
\includegraphics[width=7cm,height=5.7cm]{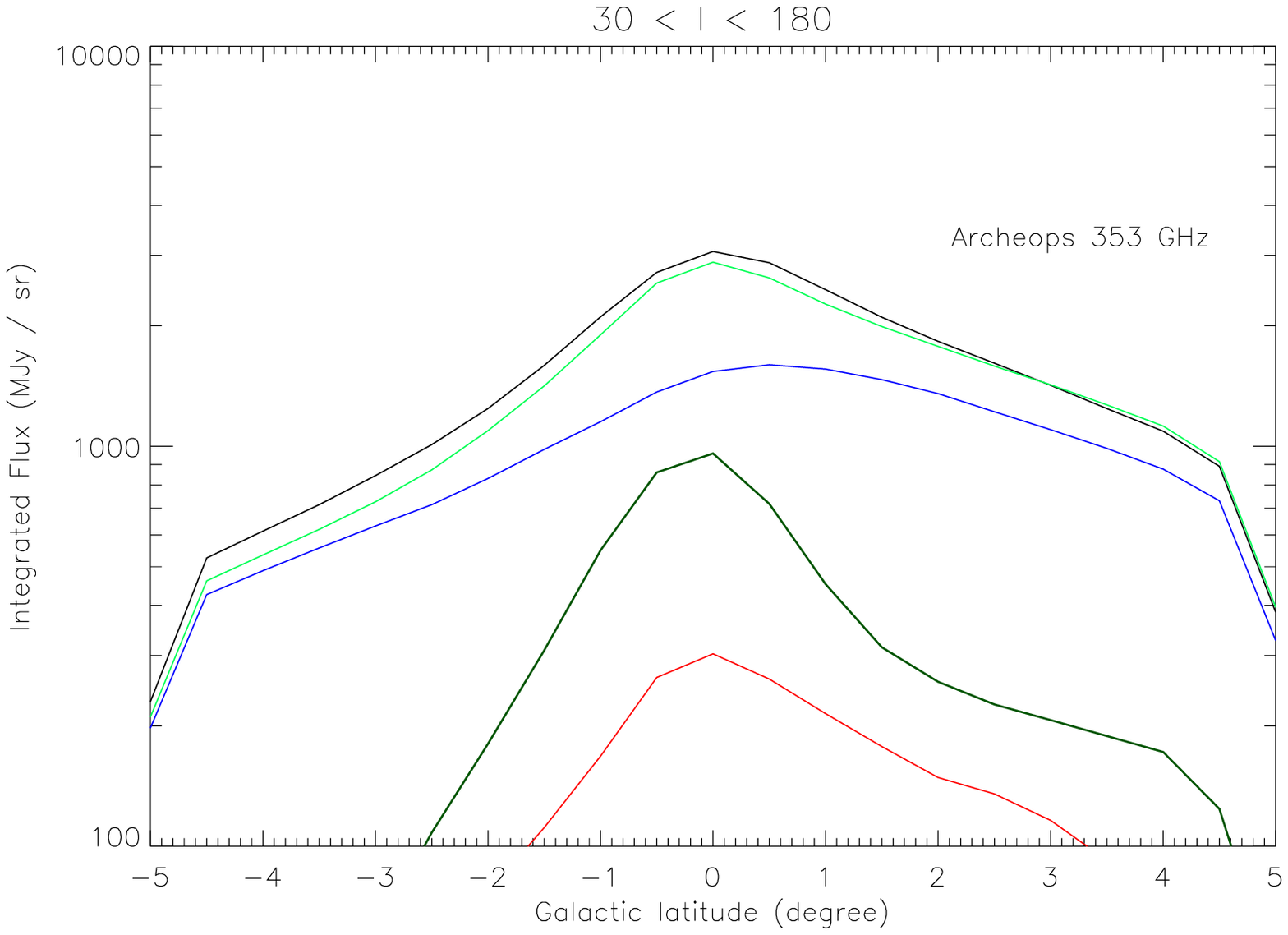}
\includegraphics[width=7cm,height=5.7cm]{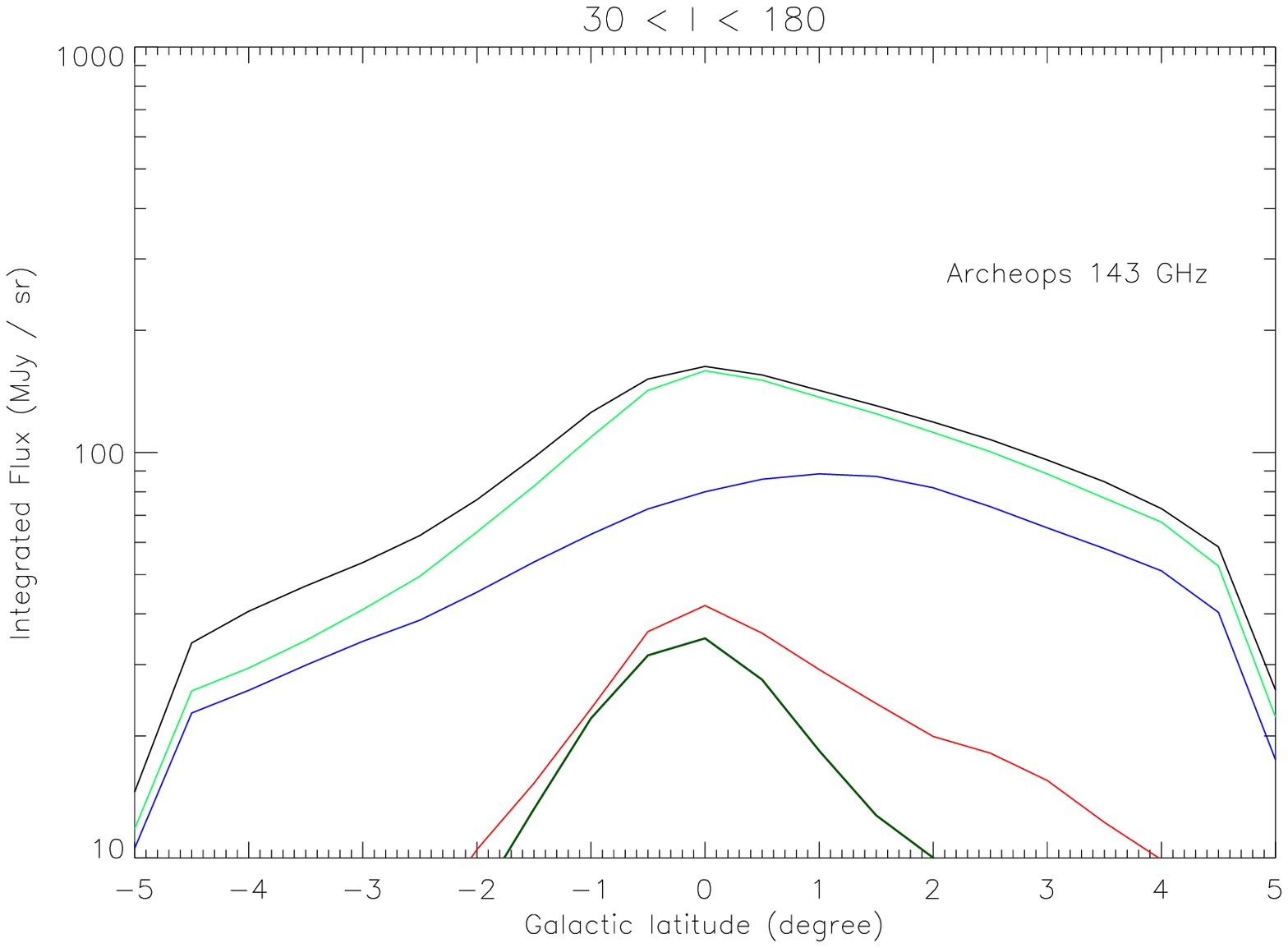}\\
\includegraphics[width=7cm,height=5.7cm]{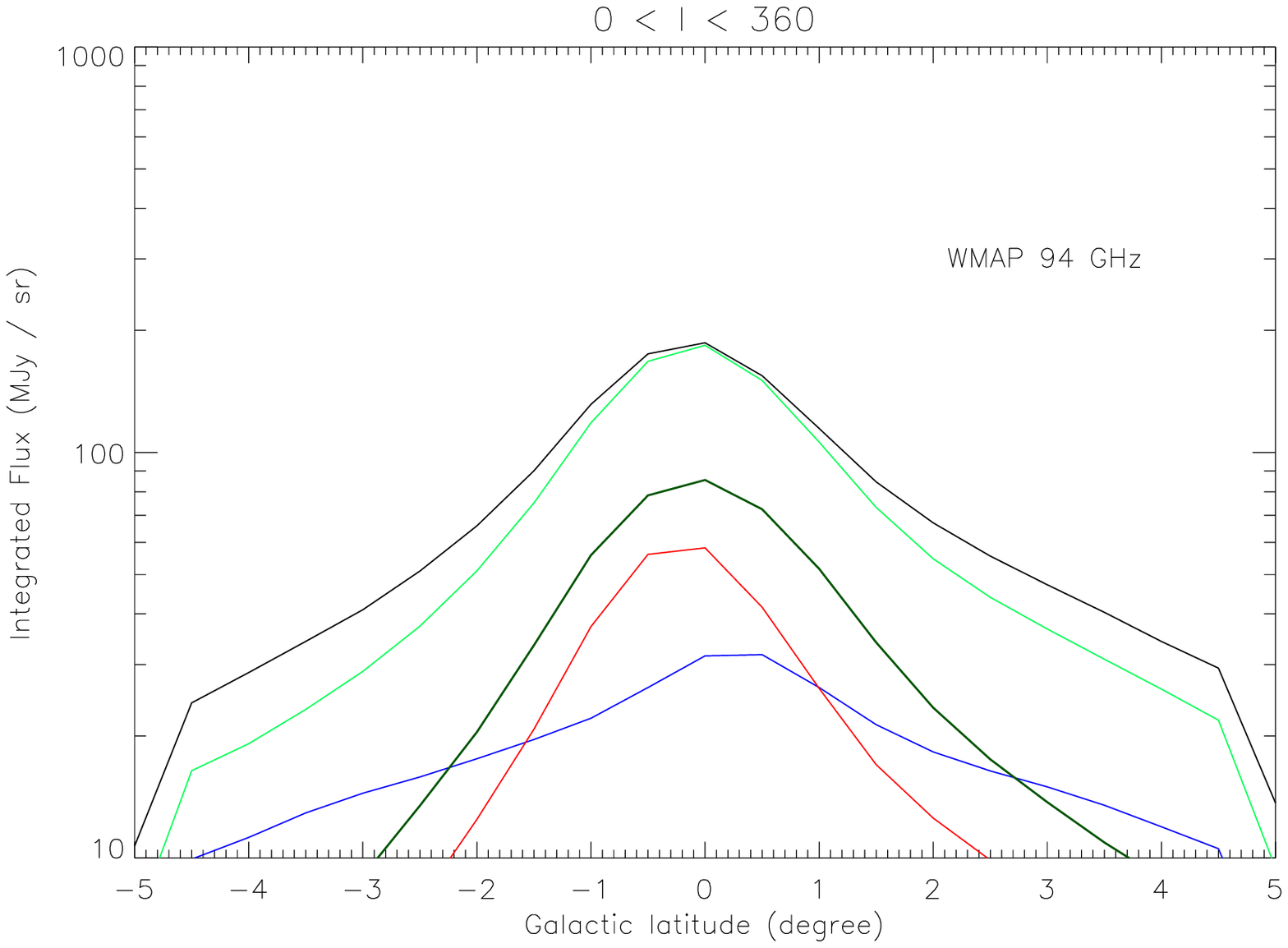}
\includegraphics[width=7cm,height=5.7cm]{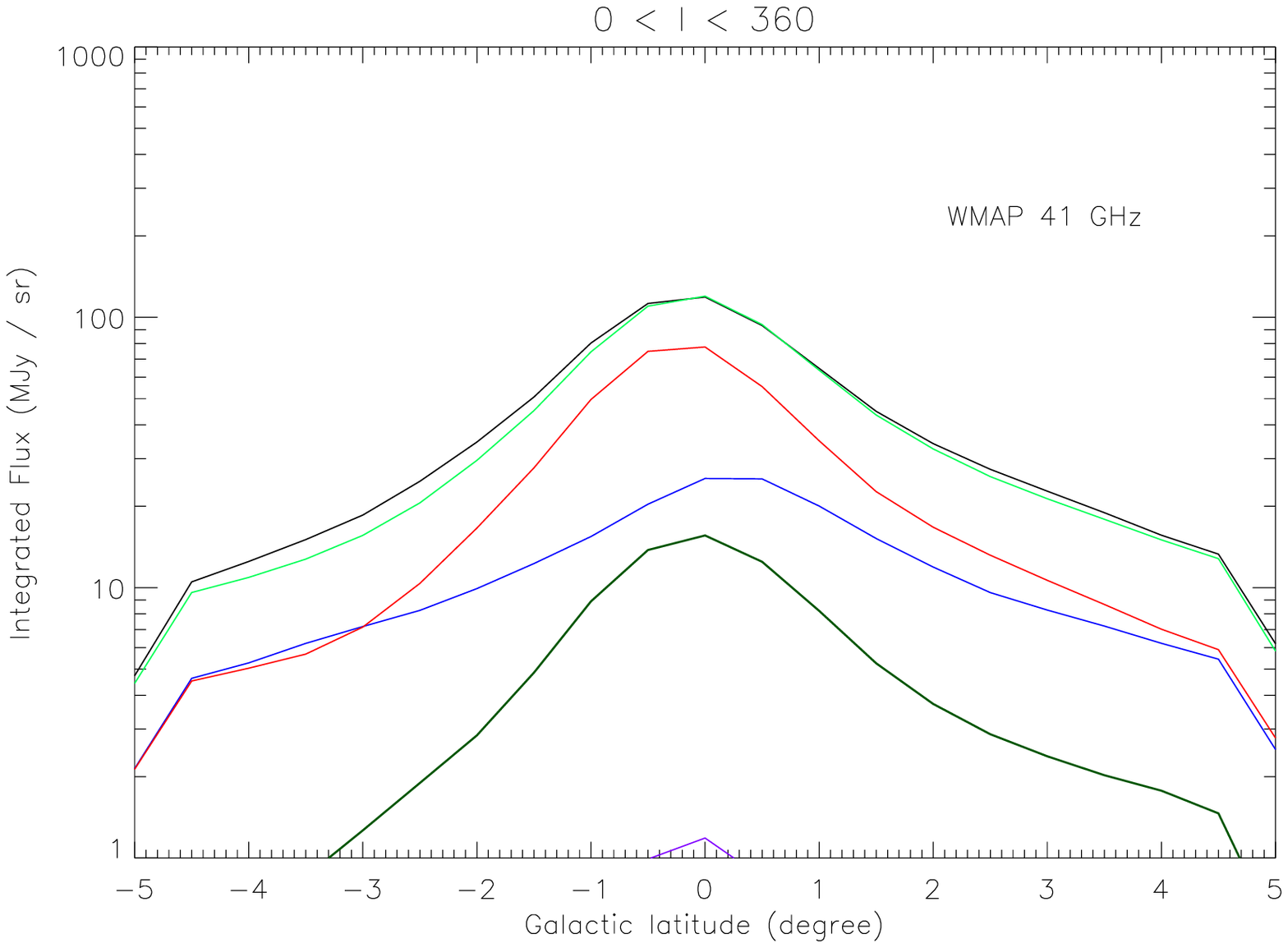}\\
   \caption{Latitude profiles (from top left, clockwise) for: IRAS 60 and 100 $\mu$m, DIRBE 140 and 240
$\mu$m, Archeops 850 and 2096 $\mu$m, WMAP W and Q bands. Black line denotes observed 
emission while light 
green line corresponds to the
fitted model. In addition, blue line is for H$_{\sc{I}}$, dark green line for H$_{\sc{2}}$, red line for H${_{\sc{II}}}^{diffuse}$ and 
magenta line
for H${_{\sc{II}}}^{compact}$.}
\end{figure*}

\noindent
Uncertainties quoted in the table have been computed by applying the bootstrap method described 
in Giard et al. (1994).

\begin{figure}
\includegraphics[width=8cm,height=8cm,angle=0]{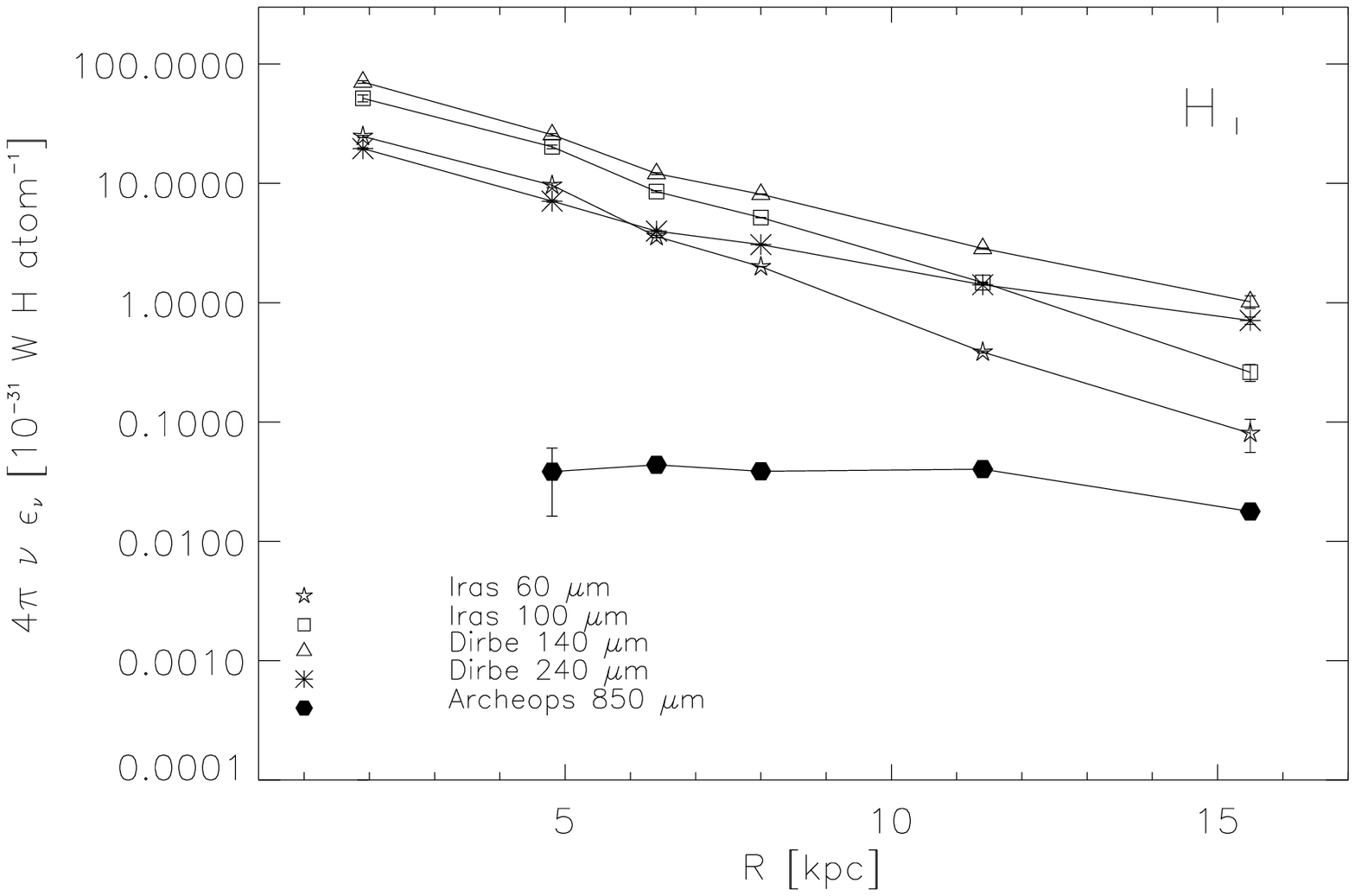}\\
\includegraphics[width=8cm,height=8cm,angle=0]{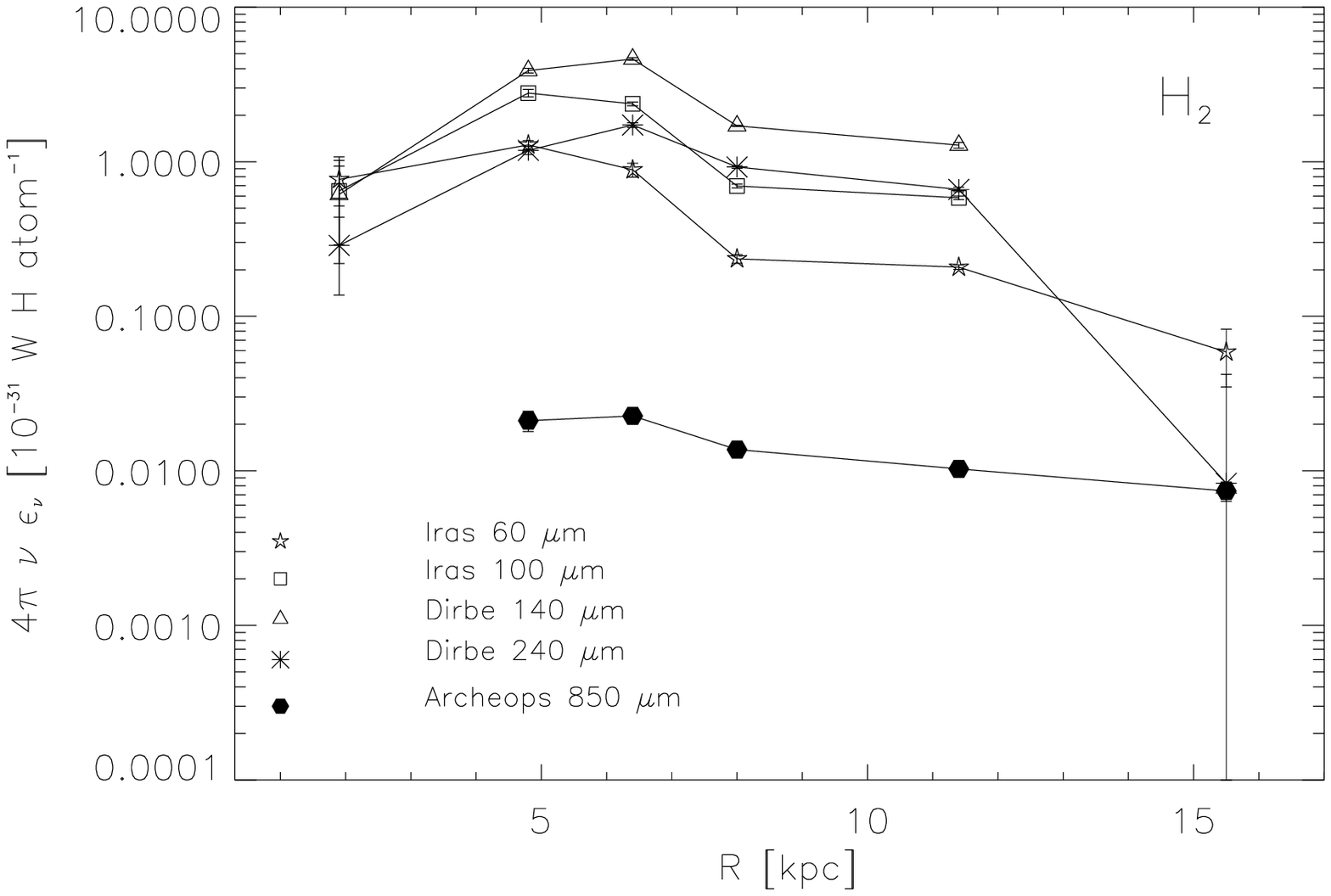}\\
\includegraphics[width=8cm,height=8cm,angle=0]{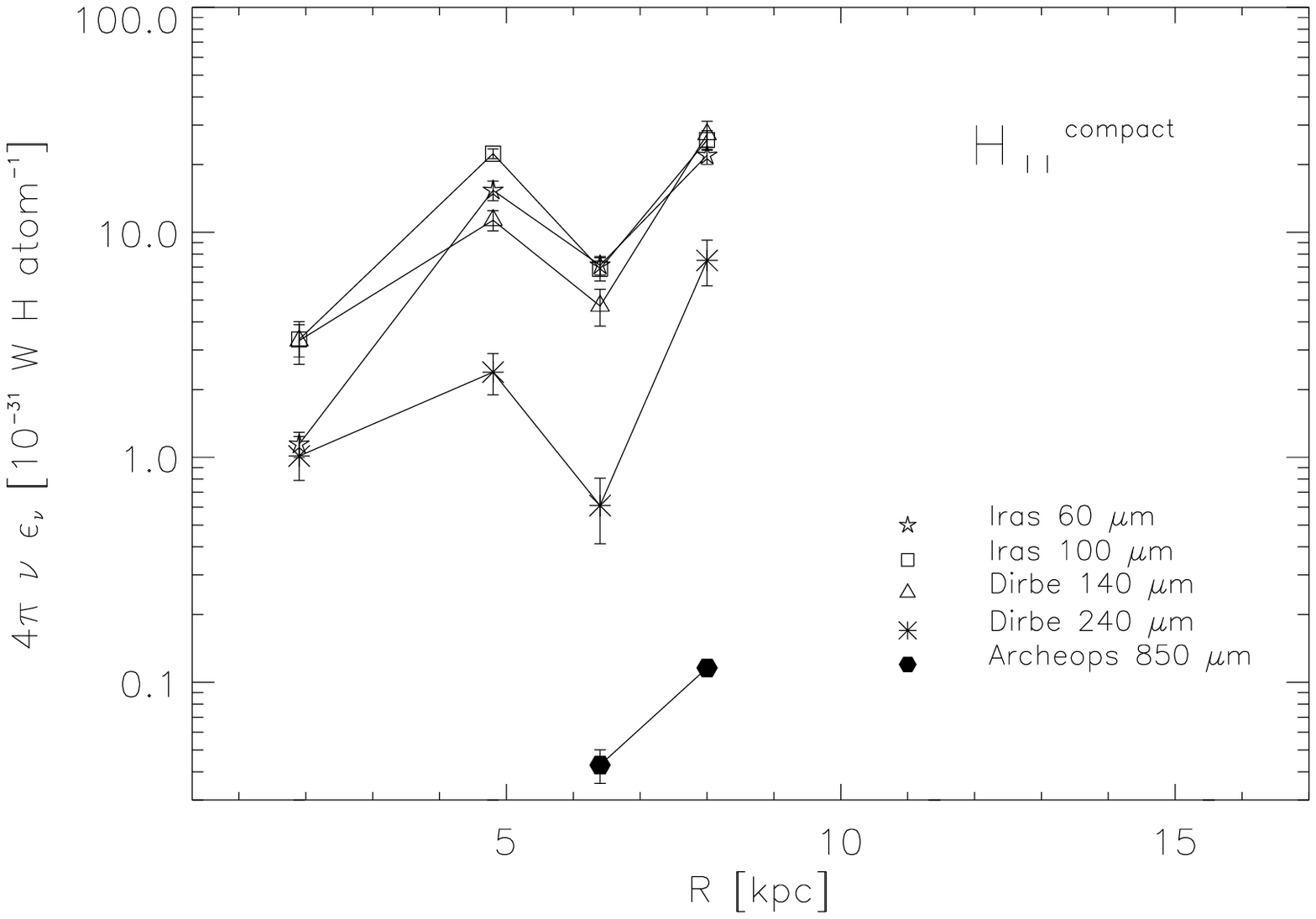}
      \caption{Galactocentric distributions of the infrared emissivities for dust associated with  
H$_{\sc{I}}$ (top left), H$_{\sc{2}}$ (top right) and H${_{\sc{II}}}^{compact}$ (bottom center). Archeops 
results at 850 $\mu$m have been obtained for the longitude interval 30$^{\circ} <$ l $<$ 180$^{\circ}$.}
\end{figure}

\subsection{Results and discussion}

\noindent
Fig.~4 and Fig.~5 show longitude profiles of the observed and
modeled intensities as well as the respective contributions to the calculated profiles from 
H$_{\sc{I}}$, H$_{\sc{2}}$, diffuse H${_{\sc{II}}}^{diffuse}$ and H${_{\sc{II}}}^{compact}$. Fig.~6 illustrates the 
corresponding latitude profiles. The 
fitted region in the longitude profiles, at each frequency, covers the latitude range $|b| <$ 5$^{\circ}$, due to the limited
sky coverage of the CO observations. We also excluded the longitude range $|l| <$
5$^{\circ}$ and the corresponding region in the anticenter due to the lack of kinematic resolution in these directions.
Although we could have corrected for this effect by, for instance, constraining the radial velocity interval, following the method 
implemented by Bertsch et al. (1993), we decided to ignore such a problem in the current analysis and 
to investigate the Galactic center region in detail in a forthcoming paper. \\

The longitude profiles show that the model fits are satisfactorily good at all wavelengths with the 
exception of a few peaks at 60 $\mu$m 
(which are not correctly 
reproduced) and some overshooting in the WMAP bands. The peaks are likely H$_{\sc{II}}$ regions and 
the model fails in this case because the H$_{\sc{II}}$ region catalog of Paladini et al. 
is complete only for fairly bright sources (see Section 3.4.2). In addition, as discussed in 
Section 3.4.2, the angular resolution of the WMAP free-free map does not allow one to 
resolve discrete sources. The overshooting testifies to an excess of emission 
and this can be explained by the fact that WMAP channels are mostly dominated by thermal bremstrahlung  
instead of dust emission (see also Fig.~8 and related discussion). 

Similarly, the latitude profiles show that the model reproduces quite well the 
observed emission. Moreover, one can see that the emission between 60 and 2096 $\mu$m is 
dominated by dust associated with H$_{\sc{I}}$ although at 60 $\mu$m this is largely contributed 
by dust associated with the ionized gas. Remarkably, the H$_{\sc{II}}$ region contribution is 
clearly visible in the IRAS 60-$\mu$m profile but almost disappears in the corresponding plot 
for WMAP at 41 GHz despite the fact that the emission, at this wavelength, is mostly thermal radiation from the ionized gas. 
As discussed previously, this is an effect due to the angular resolution: IRAS resolving power matches 
the average angular size of cataloged H$_{\sc{II}}$ regions while WMAP resolution is 10 times lower.

Fig.~7 shows the Galactocentric distribution of dust emissivities associated with H$_{\sc{I}}$, H$_{\sc{2}}$ and 
H$_{\sc{II}}$ regions. The case of diffuse H${_{\sc{II}}}$ is not plotted since for that component we do not 
have any 3d-spatial information. The emissivities for dust associated with H${_{\sc{I}}}$ decreases strongly 
with increasing Galactocentric distance at all wavelengths with the exception of 
850 $\mu$m  which shows a less pronounced trend. Such behaviour appears both inside (R $<$ R$_{\odot}$) and outside 
(R $>$ R$_{\odot}$) the solar circle. On the contrary, the 
radial distribution for the 
emissivities of dust associated with H$_{\sc{2}}$ shows a bump, at all $\lambda$, at R $\sim$ 5-6 kpc. Similarly, 
the distribution for dust 
associated with H$_{\sc{II}}$ regions presents a first pronounced bump  at R $\sim$ 5 kpc and a second at R $\sim$ 8 kpc. 
In the case of H$_{\sc{I}}$, if we combine this information with the fact that the intensity  
of the ISRF scales with Galactocentric distance in a comparable way (Mathis et al. 1983), we 
can conclude that dust mixed with atomic hydrogen is mostly heated by the general radiation 
field, in agreement with previous findings by Bloemen et al. (1990) and Sodroski et al. (1997). 
The behaviour of dust associated with molecular hydrogen suggests instead a correlation 
with the intense star formation activity taking place in the molecular ring, i.e.   
dust associated with H$_{\sc{2}}$ appears to be heated also (and perhaps by a large fraction) by massive O and B stars still embedded in the parent molecular 
clouds rather by only the general radiation field . Sodroski et al. (1997) reached a similar conclusion while Blomen et al. (1990) 
did not find any significant correlation between the radial distribution of dust emissivity 
associated with H$_{\sc{2}}$ and the molecular ring.  
As for H$_{\sc{II}}$ regions, the derived radial distribution of emissivities is 
fully consistent with the radial distribution of sources found in Paladini et al. (2004; see their Fig.~3). \\

The recovered emissivity coefficients allow one to compute the 60 $\mu$m/100 $\mu$m ratio 
in each Galactocentric bin, as shown in Table~3. The ratio is fairly constant for dust mixed with molecular gas with the 
exception 
of the first ring which is however characterized by a significant error. In the case of dust mixed with atomic gas though, 
we find a significant decreasing trend with Galactocentric radius although, within the solar circle (R $<$ 8.5 kpc), the 
gradient appears to be quite small. This second result partly contradicts Bloemen et al.'s finding that 
the 60 $\mu$m/100 $\mu$m ratio is nearly constant up to R $\sim$ 17 kpc. Nevertheless, when we take the mean values, we obtain 
60 $\mu$m/100 $\mu$m ratios of $\sim$ 0.22 and 0.23 for, respectively, dust associated with atomic and molecular 
gas{\footnote{The latter has been computed without the value for the first ring.}} in agreement with Bloemen et al. (who 
report average values of $\sim$ 0.27 and 0.20) and also with typical values at high latitudes (for instance, Boulanger $\&$ 
Perault (1988) derive an average value of 0.21$\pm$0.02 at $|b| <$ 10 deg).

\begin{table*}[h]
\begin{center}
\begin{tabular}{ccc}
\hline
\hline  
\\
{\em Gas phase} &  Ring  &  $\epsilon_{60\mu m}$/$\epsilon_{100\mu m}$ \\
                & (kpc)  &                                           \\
\hline
\\
H$_{\sc{I}}$              & 0.1-4 &    0.28$\pm$0.02    \\
                & 4-5.6 &    0.28$\pm$0.01     \\
                & 5.6-7.2 &  0.25$\pm$0.01      \\
                & 7.2-8.9 &  0.23$\pm$0.002      \\
                & 8.9-14  &  0.16$\pm$0.003      \\
                & 14-17 &    0.14$\pm$0.06      \\
\\
H$_{\sc{2}}$        & 0.1-4 &       0.7$\pm$0.51\\
                & 4-5.6 &      0.27$\pm$0.02 \\  
                & 5.6-7.2 &    0.22$\pm$0.02  \\
                & 7.2-8.9 &    0.22$\pm$0.01   \\
                & 8.9-14  &    0.2$\pm$0.01  \\
                & 14-17 &       -----\\
\\
\hline
\hline
\end{tabular}
\caption{60 over 100 $\mu$m ratio for dust associated with H$_{\sc{I}}$ and H$_{\sc{2}}$.}
\end{center}
\end{table*}

\begin{figure*}[h]
\includegraphics[width=9cm,height=8.5cm,angle=90]{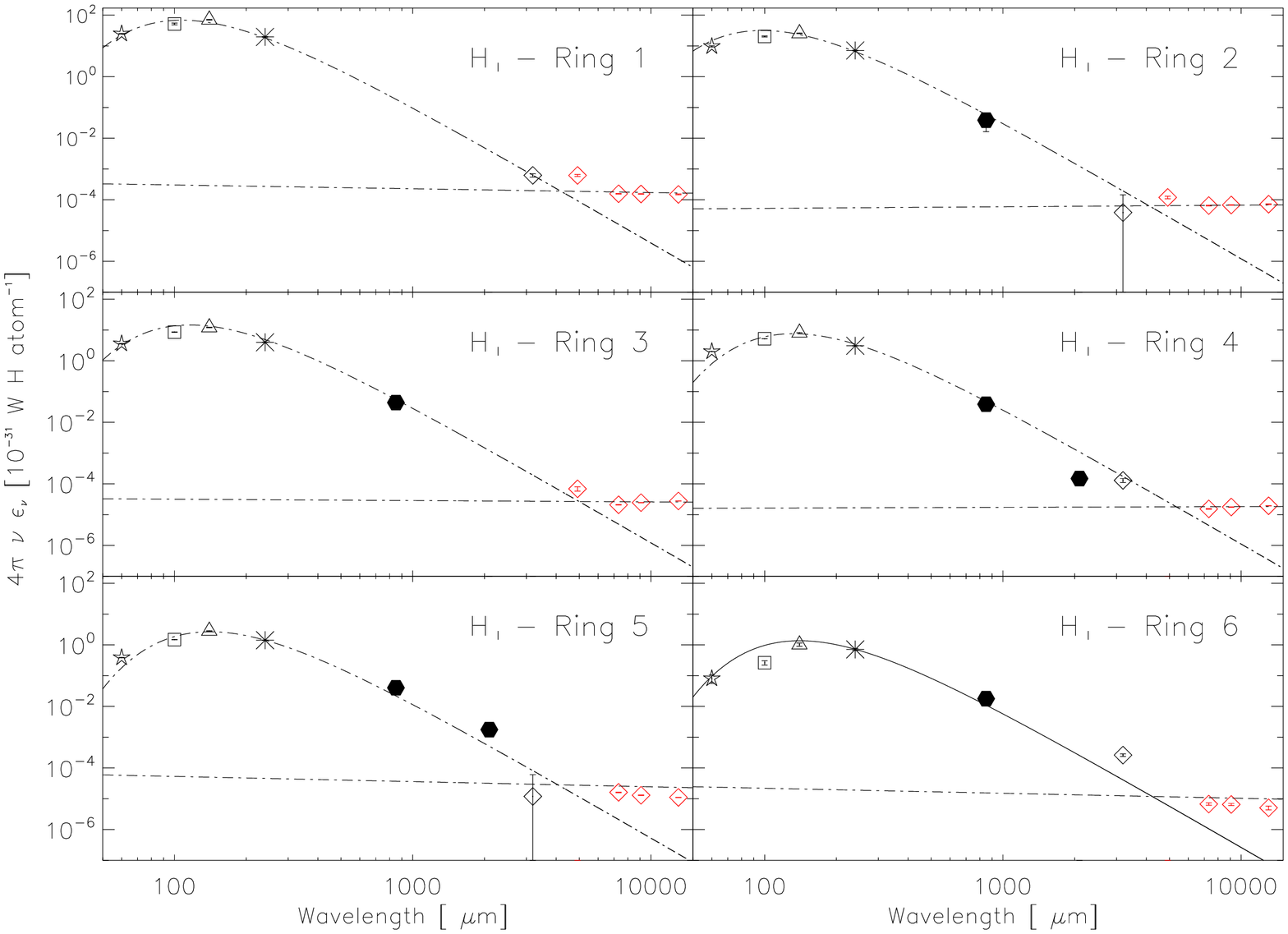}
\includegraphics[width=9cm,height=8.5cm,angle=90]{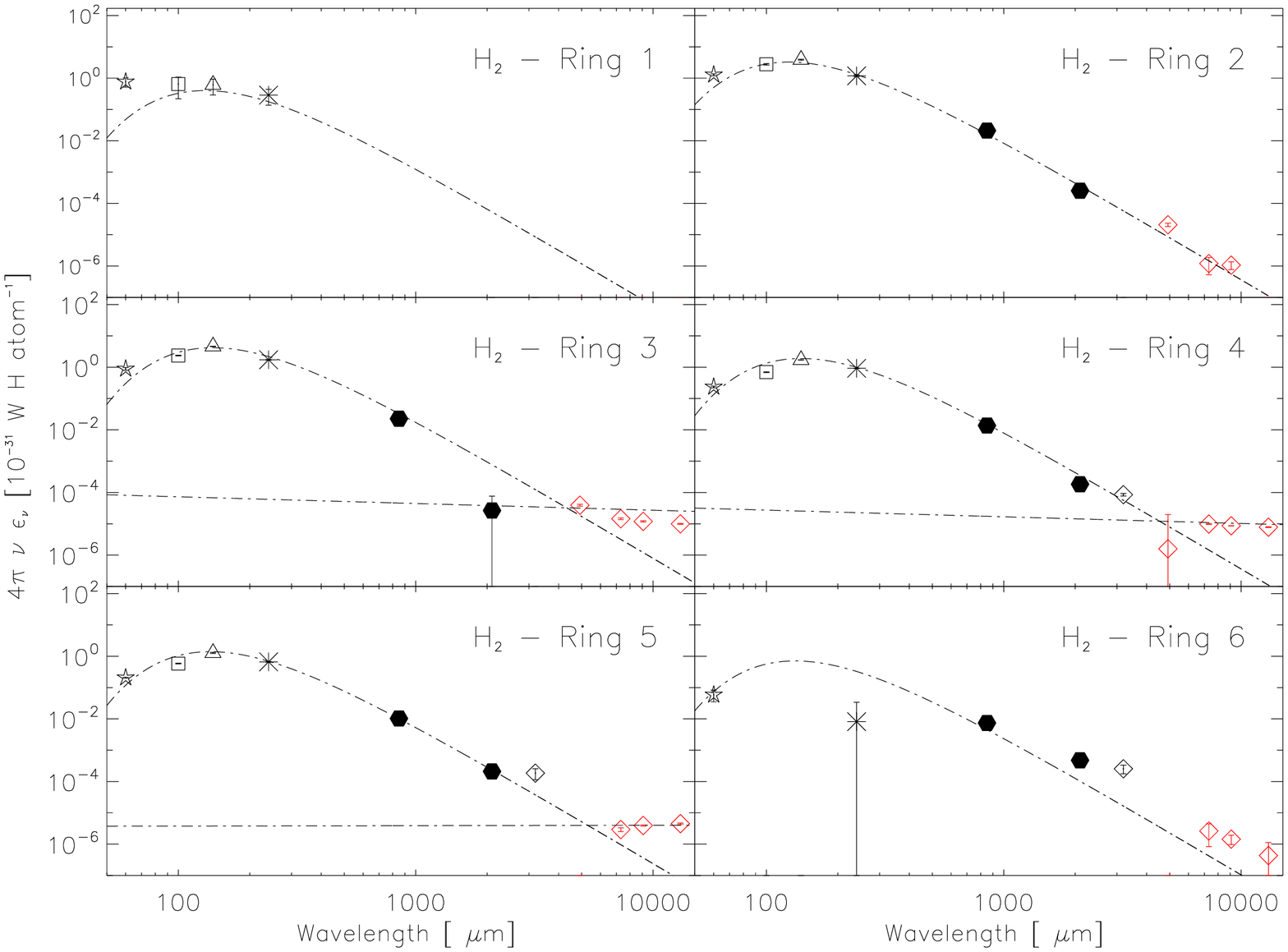}
\includegraphics[width=9.cm,height=8.5cm,angle=90]{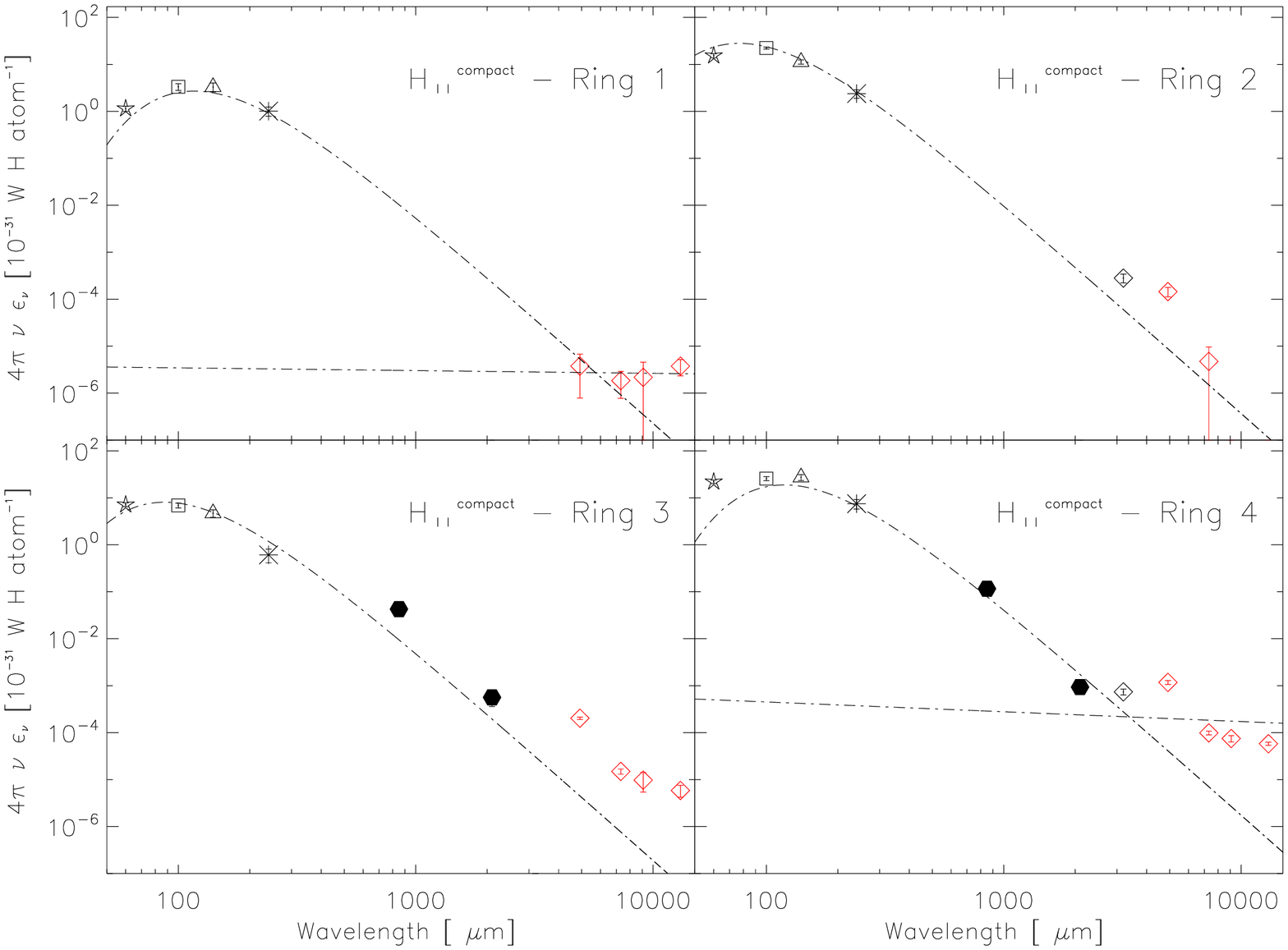}
\includegraphics[width=5.5cm,height=6.5cm,angle=90]{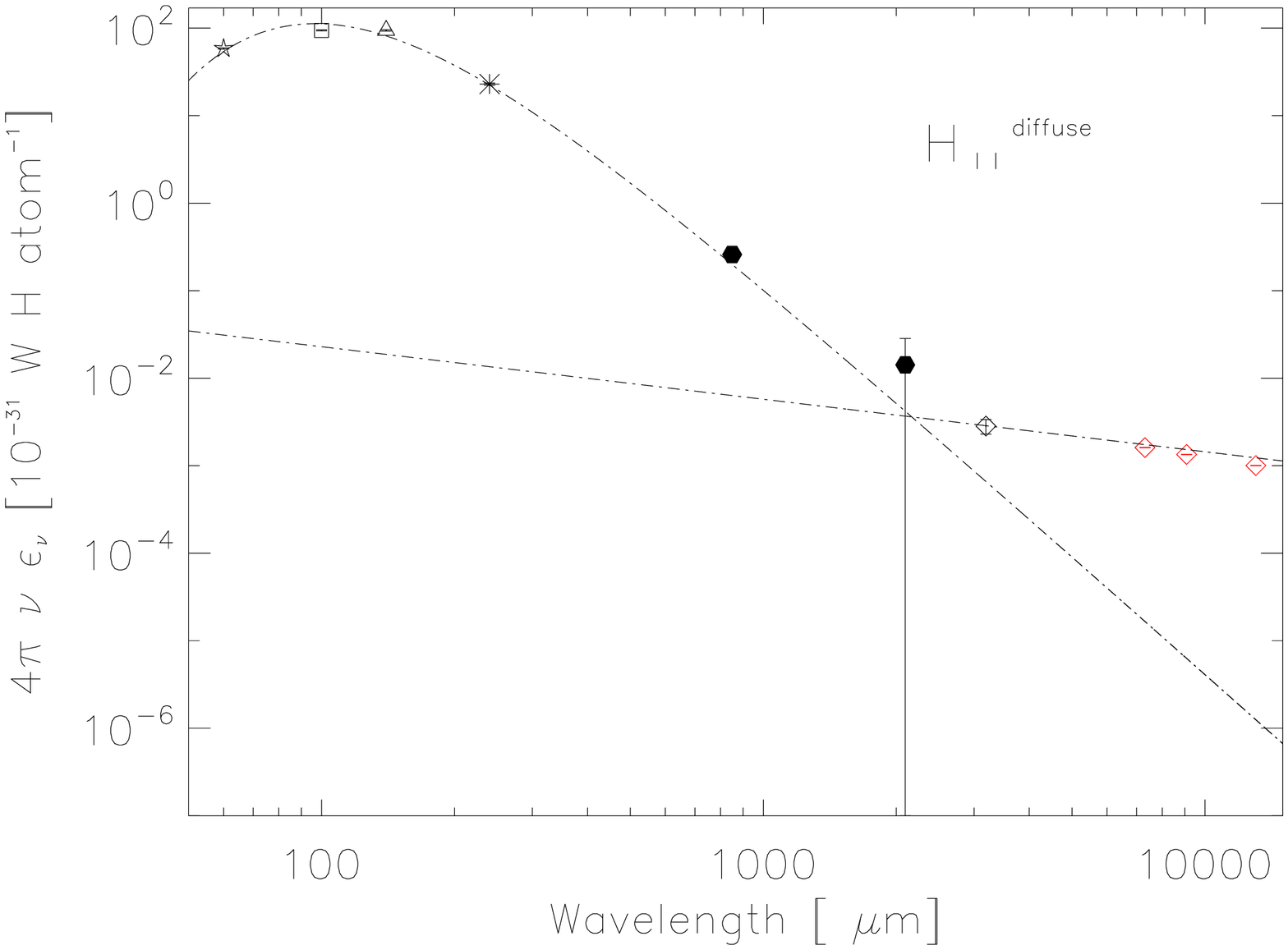}
      \caption{Derived emission spectra for dust associated with H$_{\sc{I}}$ (top left), H$_{\sc{2}}$ 
(top right), H${_{\sc{II}}}^{compact}$ (bottom left) and H${_{\sc{II}}}^{diffuse}$ (bottom right). Dashed lines illustrates 
best-fitted modified black-body and power-law functions. Symbol notation is as in 
Fig.~7: the star corresponds to IRAS 60 $\mu$m, the square to IRAS 100 $\mu$m, the triangle to DIRBE 
140 $\mu$m, the asterisk to DIRBE 
240 $\mu$m, the filled circle to Archeops 850 $\mu$m and the black diamond to WMAP W-band. In 
addition, red diamonds denote WMAP K, Ka, Q and V bands.}
\end{figure*}

\noindent
The derived IR emission spectra for dust associated with each phase of the gas 
are shown in Fig.~8. We have fitted the FIR luminosity per H mass, $4\pi \nu \epsilon_{\nu}$ (in units of  
10$^{-31}$ W (H ATOM)$^{-1}$), with a modified-blackbody, i.e.:

\begin{equation}
4\pi \nu \epsilon_{\nu} = 4\pi\lambda\epsilon_{\lambda} = 4\pi\lambda \alpha B_{T}(\lambda) \left(\frac{\lambda}{\lambda_{0}}\right)^{-\beta} 
\end{equation}

\noindent
In this expression, $\alpha$ represents the dust cross-section per H atom 
(in units of m$^{2}$/ H atom).

We have assumed a spectral emissivity index $\beta$ of 1.5.  Such a value represents a compromise
between the results obtained by Dupac et al.  (2003) for M17, i.e. 1.3
$\le \beta \le$ 1.8 in the temperature range 35-15 K, and the fact that
$\beta$=1.7 is usually adopted in the Solar neighborhood (Lagache et
al.  2003).  This flat spectral index is indicative of the existence
of a millimeter excess with respect to the $\beta=2$ behavior expected
from several grain models.  Such an excess was observed for the first
time in analyzing FIRAS data at $\lambda >$ 500 $\mu$m (Wright et al.
1991).  A similar finding is also supported by observation of external
galaxies as shown by Galliano et al.  (2005).  Finkbeiner et al.
(1999) attribute this excess to a very low temperature (9 K) emission
component in the diffuse ISM. Alternatively, the excess could be
attributed to a temperature dependence of the dust submm emissivity
spectral index (Meny et al. 2006)

By fitting the spectra with eq.~(13), we derive estimates
of the dust temperatures for every gas phase and for each radial
interval (see Table~4).  The following average temperatures have been
obtained: T$_{d}$(H$_{I}$)=19.86$\pm$2.84 K,
T$_{d}$(H$_{2}$)=19.23$\pm$2.18\footnote{This value is computed by averaging
ring 2 to 5. Ring 1 and 6 have been excluded given the significant
uncertainty.} K, T$_{d}$(H${_{\sc{II}}}^{diffuse}$)=26.7$\pm$0.1 K and T$_{d}$(H${_{\sc{II}}}^{compact}$)=27.87$\pm$2.89 K.\\

Clearly, the WMAP points (with the exception of W band) cannot be
fitted by means of a modified black-body.  Likely, this wavelength
range is dominated by a mixture of thermal bremstrahlung (free-free),
synchrotron emission and possibly (see Section 3.4) spinning dust rather
than thermal radiation from {\em stable} dust in the diffuse ISM. This combination
of radiative processes is not easy to model since the 3d-templates are
not available and the synchroton spectral index varies spatially.
Therefore, we fit the WMAP V to K bands with a power-law of the form
$\alpha^{'}$ $\lambda^{\beta^{'}}$ where $\alpha^{'}$ is a constant of
normalization and $\beta^{'}$ the spectral index for such a mixture of
emission mechanisms.  The power-law fit appears to represent
satisfactorily the emissivity coefficients derived through the
inversion technique. The best-fit $\alpha^{'}$ and ${\beta^{'}}$ 
values are shown in Table~4.

An additional interesting result of the fitting procedure is
represented by the analysis of the Archeops point at 2096 $\mu$m as
well as by the WMAP W-band value.  In the outer Galaxy (Fig.~8, Ring 5 
and 6) these points show a tendancy to lie above the fitted modified
black body.  This indicates that the galactic millimeter excess emission might
be comparatively stronger towards the outer regions of the Galaxy.  A
similar finding was reported by Bourdin et al.  (2002) also based on the
use of the Archeops data.  It is important to emphasize here that this 
result cannot be an artifact of our data analysis techniques.  Such a
spurious effect might be a matter of concern for the Archeops data set:
the maps produced by the Trapani and Kiruna flights (Delabrouille, J.
$\&$ Filliatre, P., 2004) have in fact been processed in a slightly
different way, the Trapani map being significantly more filtered than
the Kiruna map to correct for systematics.  However, since such
filtering affects the larger rather than the smaller scales, it would
tend to decrease rather than increase the observed excess.  At the same
time, we point out that this result has been obtained by fixing $\beta$
to 1.5.  A $\beta$ of 2 would enhance the excess even more.\\

The emissivity coefficients can also be combined with information
on the mass distribution of the gas to compute FIR luminosities. 
In particular, if we denote with $i$ a given phase of the gas, the total 
FIR luminosity for this phase is obtained according to:

\begin{equation}
L_{FIR,i} = M_{i} \int_{\lambda_{min}}^{\lambda_{max}} {4\pi\lambda \epsilon_{i,\lambda} d\lambda} \hskip 1.8 truecm [10^{8} L_{\odot}]
\end{equation}

\noindent
where M$_{i}$ is the mass of the gas (expressed in 10$^{8}$ M$_{\odot}$), 
4$\pi\nu\epsilon_{\lambda}$ is the FIR luminosity per H mass (in L$_{\odot}$/M$_{\odot}$ units) and 
the integral is taken over the interval [$\lambda_{min}$,$\lambda_{max}$] = [60 $\mu$m, 2096 $\mu$m]. 
In writing eq.~(14) we use again the fact that 4$\pi\nu\epsilon_{i,\nu}$ = 4$\pi\lambda\epsilon_{i,\lambda}$. 
As well as for the total Galaxy, FIR luminosities have been computed, for each phase of the gas, 
in each ring. In this case, the quantity M$_{i}$ represents the mass of the gas 
in the considered Galactocentric interval. 
Masses for the atomic and molecular hydrogen
have been derived by integrating, for each ring, the surface densities
quoted by Binney $\&$ Merrifield (1998).  As far as the ionized gas is concerned, we
have made direct use of the estimate of the total mass reported by Westerhout (1958) for an
effective electron density n$_{e}$=10 cm$^{-3}$.  Remarkably, the value
provided by Westerhout is a factor $\sim$ 40 lower than the one given
in Ferriere (2001). The two values have been obtained in a different way: the 
Westerhout's estimate is derived from
free-free emission observations which are proportional to the emission measure (EM) 
while the Ferriere's value is based on the Cordes et
al.  (1991) model which provides
a direct measure of the electron column density (DM)  
(Katia Ferriere, {\em{private communication}}). 
The fact that our Galaxy contains several thermal components which are characterized by 
different clumping factors, i.e. different relations between DM and EM, 
explains the discrepancy in the two estimates.\\
The resulting FIR luminosities are given in Table~5.\\

Finally, we have computed the contribution of each component to the 
global SED. Fig.~9 shows that the dominant contribution at all wavelengths $>$ 60 $\mu$m 
is due to dust associated with atomic gas. However, at $\lambda$ $\sim$ 60 $\mu$m the 
emission is largely contributed by dust in H${_{\sc{II}}}$ regions.

\begin{table*}[h]
\begin{center}
\begin{tabular}{cccccc}
\hline
\hline
\\
{\em Gas phase} &  Ring  &  $\alpha$ &  T &  $\alpha^{\prime}$ & $\beta^{\prime}$\\
                & (kpc)  &   (m$^{2}$/ H atom)      & (K)      &  (10$^{-31}$ W H atom$^{-1}$)       &     \\
\hline
\\
H$_{\sc{I}}$              & 0.1-4 &   0.01$\pm$0.0003 & 22.2$\pm$0.2   &  0.0005$\pm$9.0e-05  &  -0.12$\pm$0.002     \\
                & 4-5.6 &   0.002$\pm$0.0002 & 22.9$\pm$0.5            & 4.1e-05$\pm$4.1e-06 &   0.05$\pm$0.03 \\
                & 5.6-7.2 &  0.001$\pm$0.0004 & 21.1$\pm$1.01      & 3.9e-05$\pm$2.2e-07 &  -0.04$\pm$0.0003\\
                & 7.2-8.9 &  0.001$\pm$0.0005 & 19.8$\pm$1.4      & 1.5e-05$\pm$3.1e-07  &  0.02$\pm$ 0.003 \\
                & 8.9-14  &   0.0008$\pm$0.0008 & 17.8$\pm$2.9    & 0.0001$\pm$5.2e-06  &  -0.17$\pm$0.0002 \\
                & 14-17 &     0.0007$\pm$0.002 & 15.4$\pm$7.1     &  4.6e-05$\pm$1.6e-05  &  -0.16$\pm$ 0.002 \\
\\
H$_{\sc{2}}$        & 0.1-4 &   6.9e-05$\pm$0.0004 &  22.1$\pm$21.2  &  ---- & ----  \\
                & 4-5.6 &  0.0004$\pm$0.0003  &  21.7$\pm$3.2    & 0.0002$\pm$4.2e-05  &  -0.25$\pm$ 0.0002 \\
                & 5.6-7.2 & 0.001$\pm$0.0007  & 18.7$\pm$2.4     & 0.0002$\pm$1.7e-05  &  -0.21$\pm$0.0002 \\
                & 7.2-8.9 & 0.0007$\pm$0.001  & 16.9$\pm$5.2     & 7.3e-05$\pm$6.1e-06  &  -0.21$\pm$0.0002 \\
                & 8.9-14  &  0.0004$\pm$0.001  & 17.4$\pm$7.2    &  ----  &  ----  \\
                & 14-17 &    0.0002$\pm$0.03 & 18.6$\pm$222.0   & ----  &  ----\\
\\
H${_{\sc{II}}}^{compact}$ & 0.1-4 &  0.0002$\pm$0.0002  &  23.6$\pm$3.5   & 4.5e-06$\pm$3.9e-07  & -0.06$\pm$0.003 \\
                          & 4-5.6 & 0.0005$\pm$3.8e-5  &  29.2$\pm$0.3    &  ---- &  ---- \\
                          & 5.6-7.2 &  1.5e-4$\pm$3e-5  &  30.0$\pm$0.8   & ----  &  ---- \\
                         & 7.2-8.9 &  0.0007$\pm$6.1e-5   & 28.7$\pm$0.4   & 0.0012$\pm$2.2e-05  &  -0.21$\pm$0.001   \\
                         & 8.9-14  &  ----      &  ----     \\
                         & 14-17 &    ----      &  ----    \\
\\
H${_{\sc{II}}}^{diffuse}$            &  0.1-17  & 0.004$\pm$9.5e-5    &  26.7$\pm$0.1 & 0.36$\pm$0.003 &  -0.60$\pm$1.4e-08   \\
\\
\hline
\hline
\end{tabular}
\caption{Derived temperatures for dust associated with each gas phase 
at increasing radial distance from the Galactic center.}
\end{center}
\end{table*}

\begin{figure*}
\centering   
\includegraphics[width=10cm,height=10cm,angle=0]{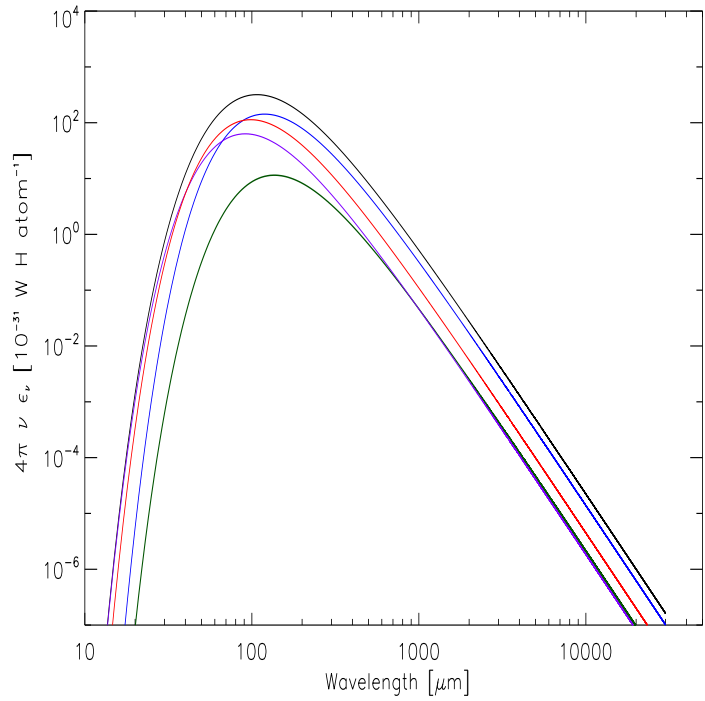}
\vspace*{0.3truecm}
\caption{Global SED for our Galaxy obtained by summing up the contributions from dust associated with atomic, molecular and 
ionized gas (black solid line). $\alpha$ and $T$ parameters are from Table~4. Also shown are the individual emission 
components: dust in atomic 
(blue line), molecular (dark green line), diffuse ionized (red line) and compact ionized (magenta line).}
\end{figure*}

\begin{table*}[h]
\begin{center}
\begin{tabular}{cccccccccc}
\hline
\hline
\\
Ring  &  ${M_{HI}}^{a}$ & $\frac{L_{FIR, H_{I}}}{M_{HI}}$ &  L$_{FIR,H_{I}}$ &  ${M_{H2}}^{a}$ & 
$\frac{L_{FIR, H_{2}}}{M_{H2}}$ &  L$_{FIR,H_{2}}$ & ${M_{HII}}^{diff,b}$ & $\frac{L_{FIR, {HII}^{diff}}}{{M_{HII}}^{diff}}$ &  
L$_{FIR, {H_{II}}^{diff}}$\\

(kpc)  &  (10$^{8}$ M$_{\odot}$) & (L$_{\odot}$/M$_{\odot}$)  & (10$^{8}$ L$_{\odot}$) & (10$^{8}$ M$_{\odot}$) 
& 
(L$_{\odot}$/M$_{\odot}$)  & (10$^{8}$ L$_{\odot}$) &  (10$^{8}$ M$_{\odot}$) & (L$_{\odot}$/M$_{\odot}$)  & 
(10$^{8}$ L$_{\odot}$) \\
\hline
\\
0.1-4    &  1.1      &  104.9   &   115.4    & 1.2     &   0.1     &  0.1      &      &   &      \\
4-5.6    &  1.8      &  25.3   &  45.6     & 2.4     &  3.7     &  8.8      &       &       &      \\
5.6-7.2  &  2.4      &  7.7   &  18.4     & 2.2     &   3.6     &  7.9      &       &       &      \\
7.2-8.9  &  3.1      &  5.2    &   16.0    &  1.1    &   1.3     &  1.4     &       &       &      \\
8.9-14   &  15.7      &  2.1     & 33.0      &  1.9    &  0.9      &  1.7      &       &       &      
\\
14-17    &  10.8      &  0.7      & 7.8      &  5.1    & 0.7       &   3.5     &     &       &      
\\
Total \hskip 0.1truecm Galaxy & 34.8   &  ----     &  263.3     &  13.9    & ----           &   23.5 
&  0.42   & 64.9     &   27.2    \\
\\
\hline
\hline
\end{tabular}
\caption{Emissivities per H mass (L$_{FIR,i}$/M$_{i}$) are computed from values of $\epsilon_{\nu}$ 
quoted in Table~2 by using the following conversion factors: 1.92 at 60 $\mu$m, 1.15 at 100 $\mu$m, 
0.82 at 140 $\mu$m, 0.48 at 240 $\mu$m, 0.13 at 850 $\mu$m and 0.05 at 2096 $\mu$m.\newline
$^{a}$ Based on surface density diagram in Binney $\&$ Merrifield (1998). \newline
$^{b}$ From Westerhout (1958). Effective electron density n$_{eff}$ = 10 cm$^{-3}$ .}
\end{center}
\end{table*}

\hskip 0.1truecm
\section{Conclusions}

\noindent
By using an inversion technique, we have derived the radial distribution of  
dust properties in our Galaxy in the wavelength range 60 to 2096 $\mu$m. 
In particular, we have obtained emissivity coefficients for dust associated with atomic, 
molecular and ionized gas. By assuming 
a modified black-body with $\beta$=1.5, we estimate that average temperatures for dust uniformly mixed 
with H$_{\sc{I}}$, H$_{\sc{2}}$ and H$_{\sc{II}}$ are of order of 19.8, 19.2 and 26.7 K respectively. It is important to emphasize that 
the derivation, through inversion techniques, of dust emissivity coefficients 
associated with the molecular and ionized phases 
may be affected by a mutual contamination due to the spatial correlation that exists between 
these phases. In turn, such contamination would affect the derived physical quantities such as 
temperature. As noted 
by Sodroski et al. (1997), only the use of RRLs to trace the ionized gas phase would allow one to 
circumvent this problem. 
From the radial distribution of the emissivity coefficients associated 
with each gas phase, we find evidence that dust associated with the H$_{\sc{I}}$ is heated by the global ISRF. On the 
contrary, 
dust in molecular clouds appears to be heated in a significant way by young massive stars still embedded in their parent 
clouds. Our best-fitted spectra show an excess at Archeops 2096 $\mu$m and WMAP 
94 GHz. Such an excess, also found in FIRAS data, can be interpreted as due to a very cold dust component 
in the diffuse ISM or to a temperature dependence of the spectral emissivity index at submm 
wavelengths. In addition, The long-wavelength ($\lambda \ge$  4900 $\mu$m) ) part of the spectra for the H$_{\sc{I}}$ 
and H$_{\sc{2}}$ associated dust components appears to be dominated by a mixture of thermal bremstrahlung, synchrotron 
emission and likely spinning dust. The possibility of a significant contribution due to spinning dust will be 
analyzed in a forthcoming paper by means of auxiliary data at even longer wavelengths ($\lambda >$ 13000 $\mu$m). 
As shown by recent studies (i.e. Watson et al., 2005), the peak of anomalous 
emission is expected to be at 15000 $\mu$m.\\
The derived emissivity coefficients also allow us to compute FIR luminosities. For these, 
the dominant contribution ($\sim$ 80$\%$) results to be provided by dust associated with atomic gas. 

\begin{acknowledgements}
      Part of this work was supported by the Marie Curie Fellowhsip 
      number EIF-502125. RP warmly thanks Katia Ferriere and Jay Lockman 
for interesting discussion on the Warm Ionized Gas. The authors also wish to 
thank an anonymous referee for a careful reading of the manuscript and for 
providing very useful comments. 
\end{acknowledgements}

\hskip 0.1truecm


\begin{thebibliography}{}

  \bibitem [2002] {benoit} Benoit et al., 2002, Astropart. Phys., 17, 101
  \bibitem[2003] {bennett1} Bennett, C. L., Bay, M., Halpern, M., Hinshaw, G., Jackson, C., 2003, ApJ, 583, 1  
  \bibitem[2003] {bennett2} Bennett, C. L., Hill, R. S., Hinshaw, G., Nolta, M. R., Odegard, N., Page, L.,  
Spergel, D. N., Weiland, J. L., et al., 2003, ApJS, 148, 97 
  \bibitem[1999] {bernard}    Bernard, J.P, Abergel, A., Ristorcelli, I., Pajot, F., Torre, J. P., et 
al., 1999, A\&A, 347, 640
  \bibitem[1993] {bertsch} Bertsch, D. L., Dame, T. M., Fichtel, C. E., Hunter, S. D., Sreekumar, P., Stacy, J. G., Thaddeus, P., 1993, ApJ, 416, 587
  \bibitem[1998] {binney} Binney, J. \& Merrifield, M., 1998, {\em Galactic Astronomy}, Princeton 
University Press, Princeton, New Jersey
  \bibitem[1990] {bloemen1} Bloemen, J. B. G. M., Deul, E. R. \& Thaddeus, P., 1990, A\&A, 233, 437
   \bibitem[1986] {bloemen2} Bloemen, J. B. G. M., Strong, A. W., Blitz, L., Cohen, R. S., Dame, T. 
M., 
et al., 1986, A\&A, 154, 25
  \bibitem[1991] {boughn} Boughn, S. P., Cheng, E. S., Cottingham, D. A., Fixsen, D. J., 1991, 
in AIP Conf. Proc. 222, {\it{After the First Three Minutes}}, ed. S. S. Holt, C. L. Bennett, V. 
Trimble, New York, p. 107
  \bibitem[1988] {boulanger1} Boulanger, F. \& Perault, M., 1988, 330, 964 
  \bibitem[1996] {boulanger2} Boulanger, F., Abergel, A., Bernard, J.-P., Burton, W. B., Desert, F.-X., Hartmann, D., Lagache, G., Puget, J.-L., 1996, A\&A, 312, 256
  \bibitem[2002] {bourdin} Bourdin, H., Boulanger, F., Bernard, J. P., Lagache, G., 2002, Ap\&SS, 281, 243  
  \bibitem[1991] {cordes0} Cordes, J. M., Ryan, M., Weisberg, J. M., Frail, D. A., Spangler, S. R., 1991, Natur., 354, 121 
  \bibitem[2002] {cordes1} Cordes, J. M. \& Lazio, T. J. W., astro-ph/0207156
  \bibitem[2003] {cordes2} Cordes,J. M. \& Lazio, T. J. W., astro-ph/0301598
  \bibitem[1986] {cox1} Cox, P., Krugel, E., Mezger, P. G., 1986, A\&A, 155, 380
  \bibitem[1988] {cox2} Cox, P., Mezger, P. G., 1988, {\it{Comets to Cosmology}}, ed. A. 
Lawrence, Springer, Berlin, Heidelberg, New York, p. 97
  \bibitem[1987] {dame1} Dame, T. M., Ungerechts, H., Cohen, R. S., deGeus, E. J., Grenier, I. A., may. J., Murphy, D. C., Nyman, L.-A., 
                  Thaddeus, P., 1987, ApJ, 322, 706
  \bibitem[2001] {dame2} Dame, T. M., Hartmann, D., Thaddeus, P., 2001, ApJ, 547, 792
  \bibitem[2004] {delabrouille} Delabrouille, J. $\&$ Filliatre, P., 2004, Ap$\&$SS, V. 290, Issue 1, p. 119
  \bibitem[1990] {desert} Desert, F.-X, Boulanger, F., Pujet, J. L., 1990, A\&A, 237, 215
  \bibitem[1990] {dickey} Dickey, J. M. $\&$ Lockman, F. J., 1990, Annu. Rev. Astron. Astrophys., 28, 215
  \bibitem[2003] {dickinson} Dickinson, C., Davies, R. D., Davis, R. J., 2003, MNRAS, 341, 1057
  \bibitem[1984] {draine1} Draine, B. T. \& Lee, H. M., 1984, ApJ, 285, 89
  \bibitem[1985] {draine2} Draine, B. T. \& Anderson, N., 1985, ApJ, 292, 494
  \bibitem[2001] {draine3} Draine, B. T. \& Li, A., 2001, ApJ, 551, 807 
  \bibitem[2001] {dupac1} Dupac, X., Giard, M., Bernard, J.-P., Lamarre, J.-M., Mény, C. et al., 2001, ApJ, 553, 604 
  \bibitem[2003] {dupac2} Dupac, X., Bernard, J.-P., Boudet, N., Giard, M., Lamarre, J.-M. et al., 2003, A\&A, 404, 11                                                    
  \bibitem[1997] {dwek} Dwek, E., Arendt, R. G., Fixsen, D. J., Sodroski, T. J., Odegard, N., 1997, ApJ, 475, 565 
  \bibitem[2001] {ferriere} Ferriere, K. M., 2001, RvMP, 73, 103 
  \bibitem[1989] {fich} Fich, M., Blitz, L. $\&$ Stark, A. A., 1989, ApJ, 342, 272
  \bibitem[1999] {finkbeiner} Finkbeiner, D. P., Davis, M. $\&$ Schlegel, D. J., 1999, ApJ, 524, 867
  \bibitem[2003] {finkbeiner2} Finkbeiner, D. P., 2003, ApJS, 146, 407 
  \bibitem[2004] {finkbeiner3} Finkbeiner, D. P., Langston, G. I. \& Minter, A. H., 2004, ApJ, 617, 350
  \bibitem[1999] {fukui} Fukui, Y., 1999, {\em{Science with the Atacama Large Millimeter Array (ALMA)}}, Associated Universities Inc.
  \bibitem[2005] {galliano} Galliano, F., Madden, S. C., Jones, A. P., Wilson, C. D., Bernard, J. P., 2005, A\&A, 434, 867
  \bibitem[1994] {giard} Giard, M., Lamarre, J. M., Pajot, F. $\&$ Serra, G., 1994, A$\&$A, 286, 
203
  \bibitem[1983] {golub} Golub, G. H. $\&$ van Loan, C. F., 1989, {\em{Matrix Computation}}, 2nd. ed., The John hopkins University Press 
  \bibitem[1999] {gorski} Gorski, K. M., Hivon, E. $\&$ Wandelt, B. D., 1999, in Proceedings of the 
MPA/ESO Cosmology Conference Evolution of Large-Scale Structure, ed. A. J. Banday, R. S. Sheth $\&$ L. 
Da Costa, 37
  \bibitem[1993] {hauser} Hauser, M. G., 1993, {\em back to the Galaxy}, ed. S. S. Holt \& F. Verter, New York AIP, AIP Proc., 278, 201 
  \bibitem[1978] {haynes} Haynes, R. F., Caswell, J. L. $\&$ Simons, 1978, Austr. J. Phys. 
Suppl., 45, 1 
  \bibitem[1968] {kerr} Kerr, F. J., 1968, in {\em{Stars and Stellar Systems}}, Vol. 7, {\em{Nebulae and Interstellar Matter}}, ed. B. M. 
Middlehurst and L. H. Aller, University of Chicago Press, p. 574 
  \bibitem[2000] {lagache} Lagache, G., Haffner, L. M., Reynolds, R. J., Tufte, S. L., 2000, A\&A, 354, 247
  \bibitem[2003] {lagache2} Lagache, G., A\&A, 2003, 405, 813
  \bibitem[1983] {lebrun} Lebrun, F. et al., 1983, ApJ, 274, 231
  \bibitem[1997] {li} Li, A. \& Greenberg, J. M., 1997, A\&A, 323, 566  
 \bibitem[1977] {mathis} Mathis, J. S., Rumpl, W., Nordsieck, K. H., 1977, ApJ, 217, 425
 \bibitem[1983] {mathis2} Mathis, J. S., Mezger, P. G., Panagia, N., 1983, A$\&$A, 128, 212
 \bibitem[2006] {Meny} Meny, C., Gromov, V., Boudet, N., Bernard, J.P., Paradis, D., Nayral, C. 2006, submitted to A\&A
  \bibitem[1982] {mezger} Mezger, P. G., Mathis, J. S., \& Panagia, 1982, A\&A, 105, 372
  \bibitem[2003] {mitra} Mitra, D., Berkhuijsen, E. M., Muller, P., 2003, {\it{How Does the Galaxy Work? A 
Galactic Tertulia with Don Cox and Ron Reynolds}}, ed. Alfaro, E. J., Perez, E., Franco, J., Astrophysics and 
Space Science Library, Published by Kluwer Academic Publishers, Dordrecht, The Netherlands, 2004, p.93 
  \bibitem[2005] {miville} Miville-Deschenes, M.-A. \& Lagache, G., 2005, APJSS, 157, 302
  \bibitem[1984] {neugebauer} Neugebauer, G., et al., 1984, ApJ, 278, L1
  \bibitem[2003] {paladini1} Paladini, R., Burigana, C., Davies, R. D., Maino, D., Bersanelli, 
M., Cappellini, B., Platania, P., Smoot, G., 2003, A\&A, 397, 213
  \bibitem[2004] {paladini2} Paladini, R., Davies, R. D., DeZotti, G., 2004, MNRAS, 347, 237
  \bibitem[2005] {paladini3} Paladini, R., De Zotti, G., Davies, R. D., Giard, M., 2005, MNRAS, 
360, 1545
  \bibitem[2003] {perrot} Perrot, C. A. \& Grenier, I. A., 2003, A\&A, 404, 519
  \bibitem[1995] {reach} Reach, W. T., Dwek, E., Fixsen, D. J., Hewagama, T., Mather, J. C., 1995, ApJ, 
451, 188 
  \bibitem[1991] {reynolds} Reynolds, R. J., 1991, {\it{The interstellar disk-halo connection in galaxies}}, 
IAUS, 144, 67
  \bibitem[1969] {schraml} Schraml, J. \& Mezger, P. G., 1969, ApJ, 156, 269
  \bibitem[1989] {sodroski} Sodroski, T. J., Dwek, E. \& Hauser, M. G., 1989, 336, 762
  \bibitem[1997] {sodroski2} Sodroski, T. J., Odegard, N., Arendt, R. G., et al., 1997, ApJ, 480, 173
  \bibitem[2001] {stepnik} Stepnik, B., Abergel, A., Bernard, J.-P., et al., 2001, ASPC, 243, 47  
  \bibitem[2005] {watson} Watson, R. A., Rebolo, R., Rubino-Martin, J.A., et al., 2005, ApJ, 624, 89
  \bibitem[1986] {weiland} Weiland, J. L., Blitz, L., Dwek, E., Hauser, M. G., Magnani, L., Rickard, L. J., 1986, ApJ, 306, 101
  \bibitem[1958] {westerhout} Westerhout, G., 1958, BAN, 14, 215
  \bibitem[1991] {wright} Wright, E. L., Mather, J. C., Bennett, C. L., et al., 1991, ApJ, 381, 200
  \bibitem[2004] {zubko} Zubko, V., Dwek, E. $\&$ Arendt, R. G., 2004, ApJS, 152, 211

\end{thebibliography}
\end{document}